\documentclass[usenatbib]{mn2e}

\usepackage{epsfig}
\usepackage{color}
\usepackage{lscape}
\usepackage{longtable}
\usepackage{journal_names}

\newcommand{\HI}{H\,{\sevensize{I}}} 
\newcommand{\B}{$B$}

\newcommand{\J}{$J$}
\newcommand{\HH}{$H$}
\newcommand{\K}{$K$}
\newcommand{\Ks}{$K_{\rm s}$}

\newcommand{\Ko}{$K^{\rm o}$}

\newcommand{\ebv}{$E(B-V)$}
\newcommand{\ab}{A_{\rm B}}
\newcommand{\av}{$A_{\rm V}$}
\newcommand{\ak}{$A_{\rm K}$}
\newcommand{\hk}{$(H-K)$}
\newcommand{\jk}{$(J-K)$}
\newcommand{\jh}{$(J-H)$}
\newcommand{\hko}{$(H-K)^{\rm o}$}
\newcommand{\jko}{$(J-K)^{\rm o}$}
\newcommand{\jho}{$(J-H)^{\rm o}$}

\newcommand{\jkoc}{$(J-K)^{\rm o,c}$}

\newcommand{\f}{$f$-value}

\newcommand{\nan}{Nan\c{c}ay}
\def\approxlt{\lower.2em\hbox{$\buildrel < \over \sim$}}
\def\approxgt{\lower.2em\hbox{$\buildrel > \over \sim$}}

\def\corresp{\lower.1em\hbox{$\buildrel \wedge \over =$}}

\newcommand{\kms}{\,km\,s$^{-1}$}
\newcommand{\etal}{et~al.}

\newcommand{\eg}{{e.g.},\ }         
\newcommand{\ie}{{i.e.},\ }         
\newcommand{\cer}{colour-extinction relation}
\newcommand{\Cer}{Colour-extinction relation}
\definecolor{grey}{rgb}{0.5,0.6,0.7}

\title[{A comparative analysis of Galactic extinction}]
      {A comparative analysis of Galactic extinction at low Galactic 
        latitudes }

\author[A.C. Schr\"oder et al.]{A.C. Schr\"oder,$^{1}$\thanks{E-mail:acs@saao.ac.za}, 
 W. van Driel$^{2,3}$, R.C. Kraan-Korteweg$^{4}$ \\
$^{1}$South African Astronomical Observatory, PO Box 9, Observatory 7935, Cape Town, South Africa\\
$^{2}$GEPI, Observatoire de Paris, PSL Research University, CNRS, 5 place Jules Janssen, 92190 Meudon, France\\
$^{3}$Station de Radioastronomie de \nan, Observatoire de Paris, CNRS/INSU USR 704, Universit\'e d'Orl\'eans OSUC,\\ 
route de Souesmes, 18330 \nan, France\\
$^{4}$Department of Astronomy, University of Cape Town, Private Bag X3,\\ 
Rondebosch 7701, South Africa\\
}

\begin{document}

\date{Accepted....... ;}

\pagerange{\pageref{firstpage}--\pageref{lastpage}} \pubyear{2017}

\maketitle

\label{firstpage}

\begin{abstract}
We use near-infrared \jk -colours of bright 2MASS galaxies, measured within
a $7\arcsec$-radius aperture, to calibrate the \citet{schlegel98}
DIRBE/IRAS Galactic extinction map at low Galactic latitudes
($|b|<10\degr$). Using 3460 galaxies covering a large range in extinction
(up to \ak$=1\fm15$ or \ebv$\simeq 3\fm19$), we derive a correction factor $f
= 0.83\pm0.01$ by fitting a linear regression to the \cer , confirming that the Schlegel
et al.\ maps overestimate the extinction. We argue that
the use of only a small range in extinction (\eg \ak$<0\fm4$) increases
the uncertainty in the correction factor and may overestimate it. Our data
confirms the \citet{fitz99} extinction law for the \J - and \K -band.  We
also tested four all-sky extinction maps based on Planck satellite
data. All maps require a correction factor as well. In three cases
the application of the respective extinction correction to the galaxy colours
results in a reduced scatter in the \cer , indicating a more reliable
extinction correction. Finally, the large galaxy sample allows an analysis
of the calibration of the extinction maps as a function of Galactic
longitude and latitude. For all but one extinction map we find a marked
offset between the Galactic Centre and Anticentre region, but not with the
dipole of the Cosmic Microwave Background. Based on our analysis, we
recommend the use of the GNILC extinction map by \citeauthor{planck16b}
(2016b) with a correction factor $f = 0.86\pm0.01$.
\end{abstract}

\begin{keywords}
ISM: dust, extinction -- 
galaxies: photometry -- 
infrared: galaxies
\end{keywords}

\section[]{Introduction} \label{intro}  

Accurate photometric measurements of galaxies, like brightness or colours,
and their derived quantities, for instance stellar masses and star
formation rates, require a correction for Galactic foreground extinction.
While such a correction is usually small and often negligible at high
Galactic latitudes, it becomes more and more important at lower latitudes
and starts affecting completeness estimates and other statistical
evaluations as well.

The first, frequently used, extinction map was based on \HI\ column density
measurements and galaxy counts, assuming a correlation between gas and dust
in the interstellar medium (\citealt{burstein78}, \citealt{burstein82}).
More recently, \citet{lenz17} have published a new map based on \HI\ column
densities, albeit covering only the low-column density, high-latitude
sky.


However, since the gas-to-dust ratio is not constant everywhere and can
vary by up to a factor of two locally (\eg \citealt{burstein87}), more
accurate extinction maps, based on far-infrared (FIR) dust emission, were
introduced by \citet[][hereafter SFD]{schlegel98}, using high-resolution
all-sky maps from the DIRBE and IRAS missions.  The advantage of these maps
is that they cover all Galactic latitudes though they could not be
calibrated in the Galactic plane ($|b|<5\degr$) due to FIR confusion
problems and lack of known galaxies.

There have been various attempts to calibrate the SFD maps in general (\eg
\citealt{dutra03} with a correction factor $f = 0.75$,
\citealt{choloniewski03} with $f = 0.71$) and at low Galactic latitudes
(the so-called Zone of Avoidance, hereafter ZoA) in particular (\eg
\citealt{nagayama04} with $f = 0.67$, \citealt{schroeder07} with $f =
0.87$). In 2011, \citeauthor{schlafly11} (hereafter SF11) 
have done a more comprehensive investigation and derived a correction
factor $f = 0.86$ for the all-sky SFD maps that has been widely accepted.

Recently, the Planck mission with a more sensitive survey of the infrared
sky has produced several whole sky extinction maps, some based on the first
data release (PR1; \citealt{planck14}, \citealt{meisner15}), others on the
full mission data (PR2; \citealt{planck16a}, \citealt{planck16b}). Although
there have been various attempts in comparing the available extinction maps
(\eg \citealt{chiang19}), most comparisons were done at high latitudes or
low extinction levels.

We have therefore decided to use a comprehensive and magnitude-limited
sample of near-infrared (NIR) bright galaxies from the 2MASX catalogue at
low latitudes ($|b|<10\degr$) and high extinction levels elsewhere (\ebv $>
0\fm95$; \citealt{schroeder19}, hereafter Paper I) to investigate the
existing extinction maps using the galaxy colours. Our main aim was to derive a
correction factor that is also valid at high extinction levels and to
search for possible variations across the Galactic plane region. We have thus
determined correction factors for both the SFD map and the four Planck
mission maps.

This paper is structured as follows. We introduce the galaxy input sample
used for our analysis in Section~\ref{sample}. Section~\ref{galext}
presents the \cer\ and investigates ways to construct the cleanest possible
sample and best parameter settings. In Section~\ref{planck} we investigate
the four extinction maps based on Planck data, and in Section~\ref{maps} we
study the variation of the calibration factor across the ZoA. We summarise
our findings and give conclusions in Section~\ref{concl}. Throughout the
paper, \K\ refers to the 2MASS \Ks -band, and the superscript `o' indicates
a correction for extinction.

\section{The sample} \label{sample} 

Paper I gives a detailed description of the selection criteria for our
galaxy catalogue, but for easy reference we recapitulate the most important
criteria here.

We selected all 2MASX\footnote{See the 2MASS All-Sky Extended Source
  Catalog (XSC) as found online at
  http://irsa.ipac.caltech.edu/cgi-bin/Gator/nph-dd} objects that satisfy
the following criteria:
\begin{enumerate}
\item \Ko\,$ \le 11\fm25$ (\ie corrected for SFD Galactic extinction);
\item $|b| \le 10\degr$ or \ebv $ > 0\fm95$.
\end{enumerate}
To correct for Galactic extinction we used the SFD maps without any
correction factor so as to be comparable to the 2MASS Redshift Survey
(2MRS; \citealt{huchra12}) selection criteria.  We visually inspected all
objects using images from various passbands (optical to NIR) to distinguish
between galaxies, blended stars, artefacts and Galactic nebulae. This
resulted in 3675 plus 88 galaxies for the 2MZOA ($|b| \le 10\degr$) and
2MEBV (\ebv $ > 0\fm95 $, $|b| > 10\degr$) catalogue, respectively.

For the our investigation, we excluded four galaxies that have an unknown
galaxy classification (class 5 in Paper I) and 166 galaxies where the 2MASX
fit is centred on a superimposed star instead of the galaxy's bulge -- we
assume all the photometry of these objects to be unreliable. That results
in an input sample of 3593 galaxies. This input sample has been carefully
investigated and the resulting final sample, as detailed in the following
section, comprises 3460 galaxies.

\section{Galactic extinction correction} \label{galext}

We use the method of calculating the extinction correction factor using NIR
colours as outlined in \citet{schroeder07}. We first apply a correction for
Galactic extinction using the SFD values (given as \ebv ), $R_V = 3.1$ (or
$R_B = 4.14$) and the \citealt{fitz99} extinction law (hereafter F99, see
Table~\ref{elawtab}), resulting, for example, in a measured \mbox{$A_{\rm
    K} = 4.14 \cdot 0.087 \cdot E(B-V)$}. If we then assume that the true
extinction $\widetilde A_K$ is a constant $f$ times the measured value from
the extinction map used, \eg $\widetilde A_K = f\,A_K $, the slope $a$ of
the relation between extinction-corrected NIR colours and extinction can be
used to calculate $f$ as
\begin{equation}
f = 1 + {a \over E/A_K} \ .
\label{fofa}
\end{equation}
where $E$ is the colour excess of the respective colour. If the SFD maps
were correct, the factor would be 1.

As explained in detail below, we decided to use \jko -colours from the
7\arcsec -aperture for this investigation. In addition to the selection
criteria for the input sample mentioned in Sec.~\ref{sample}, we `cleaned'
the sample by deselecting on the 2MASX photometry Flag~4 and by excluding
galaxies with excessive photometric uncertainties (\J -band error $ \ge
0\fm13$, \K -band error $ \ge 0\fm12$), hereafter called photometric
selection criteria. We also excluded very high extinctions, setting the
limit at \ak = 1\fm15. 

\begin{table} 
\centering
\begin{minipage}{140mm}
\caption{{NIR extinction coefficients relative to the \B -band} \label{elawtab}}
\begin{tabular}{clcc}
\hline
Passband & \multicolumn{1}{c}{$\lambda$} & \citealt{fitz99} &  \citealt{cardelli89} \\
         & ($\mu$m)  &    $A_\lambda$ / $\ab$ & $A_\lambda$ / $\ab$ \\
\hline
\J    & $\phantom{1}1.25$  & 0.208   & 0.211   \\ 
\HH   & $\phantom{1}1.65$  & 0.128   & 0.136   \\ 
\Ks   & $\phantom{1}2.15$  & 0.087   & 0.085   \\ 
\hline
\end{tabular}
\end{minipage}
\end{table}

\begin{figure} 
\centering
\includegraphics[width=0.45\textwidth]{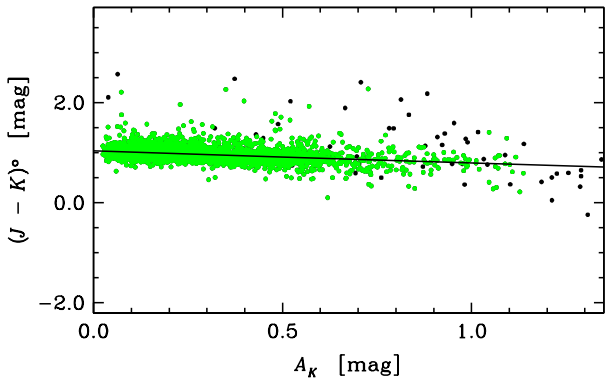}
\caption{$7\arcsec$-aperture \jko -colour versus \K -band extinction
  $A_{\rm K}$ (green dots) and linear fit for the cleaned sample. The black
  dots show the input sample before the selection criteria for the final
  sample were applied. 
}
\label{colextplot}
\end{figure}

Figure~\ref{colextplot} shows the final `cleaned' sample (green dots) and
the linear regression fit; black dots show the input sample (without the
photometric selection criteria applied and with no cut-off in extinction).
The cleaned sample comprises 3460 galaxies and the linear regression fit
gives
\[ (J-K_{\rm s})^{\rm o} = (-0.241\pm 0.012) \cdot A_{\rm K} + (1\fm036 \pm 0.004), \]
with a standard deviation or scatter of $0\fm134$\footnote{Note that
  most red outliers are caused by a combination of wrong extinction values
  and photometric problems, cf.\ Figs.~\ref{testplot}--\ref{abplot}.} . This
results in a correction factor of $f=0.83\pm0.01$. 

In the following, we will discuss the individual selection criteria in
detail.

\subsection{Colour selection}

With three passbands one can construct three colours. The \jk -colour spans
the largest wavelength range and is therefore the most reliable, as we have
confirmed in our tests. However, for internal consistency we have conducted
our investigation on the selection criteria using all three colours (for
example, it helps to understand whether an outlier is due to problems in
one passband only), but to avoid repetition, we present our results using
only the \jk -colour.

\subsection{Aperture selection }

The 2MASX catalogue offers a variety of photometric measurements, ranging
from small to large apertures, circular and elliptical apertures, isophotal
and total magnitudes, as well as peak and central surface brightnesses.
Since we are interested in the most stable colours, as little affected by
contaminating stars as possible, we prefer a smaller circular aperture and
therefore investigated those with radii 5\arcsec , 7\arcsec\ and 10\arcsec,
hereafter A5, A7 and A10. For completeness reasons, we also looked at
the isophotal magnitudes measured within the \K -band 20 mag arcsec$^{-2}$
isophotal elliptical aperture (hereafter ISO) as well as the central
surface brightness (within a radius of 5\arcsec; hereafter S5) and the peak
pixel brightness (hereafter PK). Note that the latter two do not have
errors or flags associated with them.

While we want to use as clean a sample as possible, we also want to retain
as many galaxies as possible for the subsequent investigation by
region. The two requirements are contrary, that is, we need to balance the
benefits of one against the detrimental effects of the other. To start
with, we find that up to 172 galaxies have no measurement in at least one
band per investigated aperture. This favours A7, where all galaxies have
measurements at all bands. Furthermore, bands and apertures are differently
affected by large uncertainties and flag settings (see also the following
sections). We found that 2MASX Flag 4 affects more objects the larger the
aperture, which favours A5 magnitudes. Regarding the magnitude error
distributions, we find that the isophotal magnitudes have on average $2-3$
times larger uncertainties than the aperture magnitudes, which argues
against using these magnitudes (apart from the obvious problem of increased
contamination with superimposed stars). The error distributions for the
three aperture magnitudes, on the other hand, are quite comparable.

Studying the individual linear fits, we find that the slopes depend little
on the chosen magnitude and usually agree within the 1-$\sigma$ error. The
offsets, however, are more strongly dependent on the aperture, implying a
colour growth curve across galaxies, in which isophotal colours show the
bluest colours (about $1\fm00$ for \jko ) and PK colours are reddest (about
$1\fm09$ for \jko ). The scatter (standard deviation of the sample)
decreases with aperture size (from $0\fm173$ for PK to $0\fm141$ for A5 to
$0\fm129$ for ISO), but at the same time we have about 200 galaxies fewer
for ISO than for A5 and A7 due to the increased problems with the
photometry.  Despite these variations, the $f$-value itself is fairly
independent on the aperture used and varies only by about 0.01.

Consequently, we decided to use A7, comprising the most galaxies and being
a good compromise between contamination by stars and the amount of
scatter. The arguments against isophotal colours are the smaller number of
galaxies available and larger uncertainties on average, though their
colours are truest to the galaxies. Surface brightness colours, taken from
the same aperture as the A5 colours and having no associated errors, 
are quite similar to A5 but are not available for all the galaxies. The
peak pixel brightness magnitudes are the worst with the largest scatter and
reddest colours, being strongly subjected to statistical or noise
fluctuations in a single pixel.

\subsection{2MASX photometry errors }

\begin{figure} 
\centering
\includegraphics[width=0.47\textwidth]{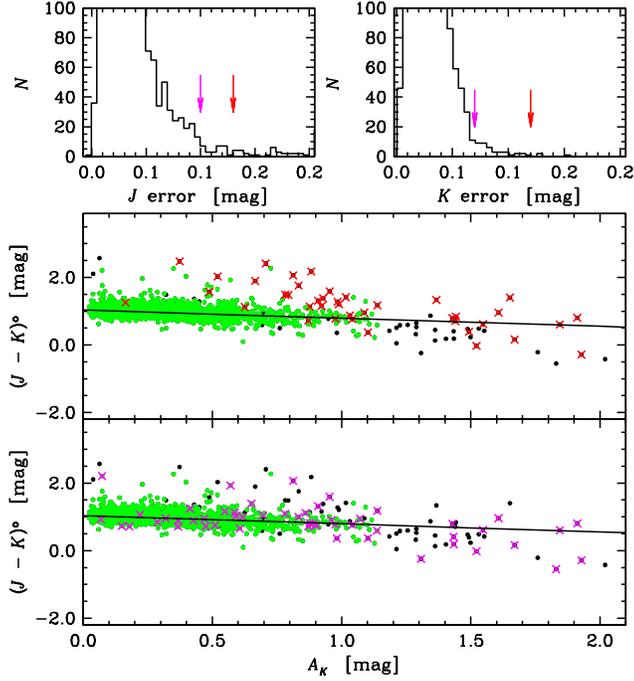}
\caption{Testing for sample selection on photometry errors for A7. Top:
  truncated histograms of the error distribution for the \J -band (left)
  and \K -band (right); the arrows indicate the two cuts tested in the
  panels below. Middle: \Cer\ for \jko ; black and green dots are the same
  as in Fig.~\ref{colextplot}; red crosses: objects with photometry errors
  above the higher cuts (see red arrows in the top panels). Bottom: Same as
  above but the magenta crosses indicate objects with photometry errors
  {\it between} the magenta and red arrows.
}
\label{testplot}
\end{figure}

Regarding photometry errors, we found that a cut at the tail end of the
error distribution removes the worst outliers in the \cer .
Figure~\ref{testplot} shows as an example the \J- and \K-band error
distributions and the \cer\ for \jk\ using A7. If we cut off at a smaller
limit (\eg\ err$^K_{\rm lim} = 0\fm07$ and err$^J_{\rm lim} = 0\fm10$,
magenta arrows), we lose too many `good' objects together with only a few
more outliers as compared to the more relaxed cut, as shown by the magenta
crosses in the bottom panel. We thus decided to use the more relaxed cuts
that clearly remove outliers without losing too many good data points, that
is, err$^K_{\rm lim} = 0\fm12$, err$^J_{\rm lim} = 0\fm13$ and err$^H_{\rm
  lim} = 0\fm10$.

\subsection{2MASX photometry flags }

The 2MASX confusion flags mostly refer to masked pixels or subtracted point
sources within the relevant aperture. We have investigated the distribution
of objects with a specific flag set in the \cer s and find that only Flag 4
(`aperture contained pixels within bright star mask') seems to indicate
outliers without affecting too many good sources. We thus exclude all
photometry where this flag was set. This affected about 15 to 180 objects
depending on aperture and colour.

\subsection{2MZOA catalogue information } \label{2mzoa}

\begin{figure} 
\centering
\includegraphics[width=0.47\textwidth]{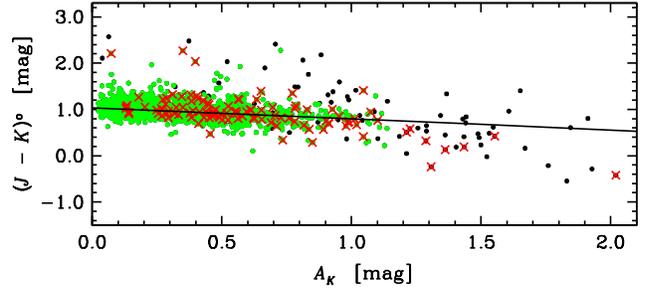}
\caption{Testing selection on extinction flags for A7, Same as
  Fig.~\ref{colextplot} with red crosses indicating objects with extinction
  flag `e' set.
}
\label{clebvplot}
\end{figure}

The 2MZOA and 2MEBV catalogues list various information regarding the
quality of data. As mentioned above, we have excluded already the four
objects which may or may not be galaxies, and all objects which are not
centred properly on the galaxy's bulge. In addition, one of the flags
indicates whether the extinction around a galaxy appears unreliable, a
question that is of strong concern to us. However, this flag is only
indicative for a strong variation in the overall area ($r\simeq 5\arcmin$)
and {\it not} of the immediate environment of the galaxy and is by no means
complete. While applying a cut on these flags removes a few outliers, too
many good sources are affected as well ($N=101$ in case of A7), see
Fig.~\ref{clebvplot}. On the other hand, the linear regression parameters
are little affected: excluding objects with this flag set changes the
$f$-value by $\le 0.003$, depending on aperture, and not at all for A7. We
therefore decided not to use this flag and instead to rely on the
conservative cut in \ak .


\subsection{Extinction cut-off }        \label{ablimit}

\begin{figure} 
\centering
\includegraphics[width=0.47\textwidth]{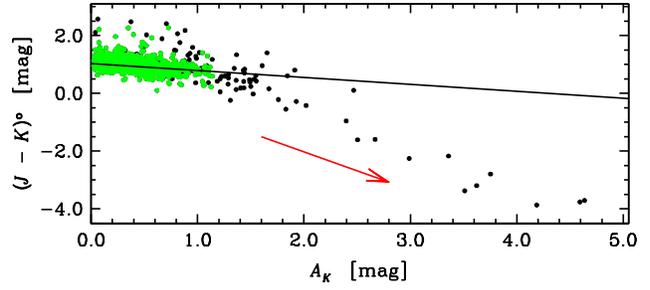}
\caption{Same as Fig.~\ref{colextplot} but showing the full extinction
  range. The reddening path is indicated with a red arrow. 
}
\label{abplot}
\end{figure}

Figure~\ref{abplot} shows the \jk\ colour distribution of the input sample
(black dots) over the full extinction range. As in Fig.~\ref{colextplot},
the regression line is based on the cleaned sample (green dots). The
colours of highly obscured galaxies become increasingly bluer, indicating a
selection bias: in very high-extinction regions, the extinction can vary
spatially very rapidly and we are more likely to find galaxies in the lower
extinction pockets. Due to the coarse spatial resolution of $6\arcmin$ of
the extinction maps, these are consequently corrected with a too-high,
average extinction value and hence appear too blue. The deviation increases
with extinction, resulting in a non-linear relation. This effect has
already been noted by, \eg \citet{vandriel09}.

\begin{figure} 
\centering
\includegraphics[width=0.45\textwidth]{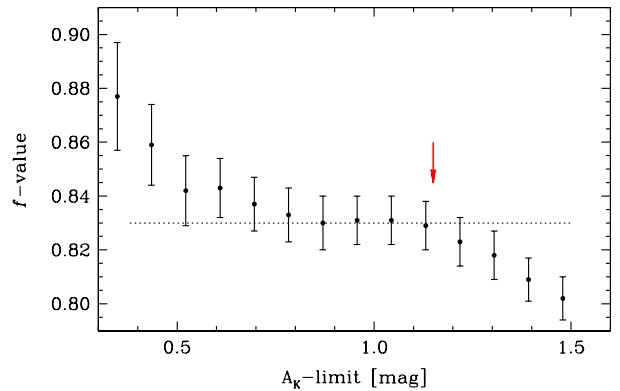}
\caption{$f$-values derived for various upper limits in \ak\ for the linear
  regression fit. The dotted line shows the plateau at $f=0.83$ and the
  arrow the chosen cut-off in \ak .
}
\label{ablimplot}
\end{figure}

To determine the best cut-off in extinction to avoid this bias for our
sample, we have used a series of upper limit values and determined each
time the $f$-value, see Fig.~\ref{ablimplot}. We find a plateau in the
range \ak $=0\fm5 - 1\fm2$. The lower $f$-values at higher extinctions are
clearly caused by the selection bias, while the higher $f$-values at the
low-extinction end are likely due to a large extrapolation beyond the small
range in extinction available for the fit and therefore being less certain,
as implied by the larger error bars.

\subsection{Other outliers }    \label{outliers}

Despite the `cleaning' of the sample some outliers remain, likely due to
insufficient star subtraction or other problems with an automated pipeline
as was used for the 2MASX catalogue. Short of re-doing the photometry by
hand, we have no simple and reliable selection criteria to get rid of
these. Instead, we tested their effect on the linear regression by removing
all data points outside a 4- and a 3-$\sigma$ envelope to the regression
line. In case of A7, 22 and 54 data points, respectively, were thus
removed. The effect is small, the $f$-value is $\sim\!0.005$ smaller in
case of a 4-$\sigma$ cut, and $0.001$ higher for the 3-$\sigma$ cut. The
other apertures give similar results. The deviations are smaller than the
error on the $f$-value and we conclude that our linear regression is robust
with respect to outliers.

\subsection{Extinction law }

As mentioned above, we have used the F99 extinction law to calculate the
NIR extinction coefficients. To understand the impact of possible
uncertainties in these coefficients on our results, we have tested the
commonly used coefficients given by \citet[][hereafter C89]{cardelli89},
see Table~\ref{elawtab}.

Theoretically, the \f\ should be independent of the colour used, and the
variations we find between the fits for the three available colours are
likely due to uncertainties in the colour measurements. These variations,
however, are noticeably larger if we use the C89 extinction
coefficients ($\Delta f = 0.17$ versus 0.06 for F99), see
Table~\ref{tabvar}, hence we prefer the F99 coefficients. Nonetheless,
while the \f s for the latter agree well for the colours \jk\ and \jh , the
\f\ for \hk\ still deviates. We can take this further and try to achieve
better agreement between the colours: since mainly the \HH -band is
affected, we varied its coefficient while leaving the others fixed. We find
very good agreement within the error if we use a coefficient of 0.130
instead of 0.128 (named `Test' in Table~\ref{tabvar}).

Note that the disagreements also increase when rounding the coefficients to
the 2-digit precision used in Paper~I. Considering the sensitivity of the
\f\ to the precision and exact value of the extinction coefficients, we
recommend to use the higher precision values to avoid any possible
systematics, even though they may be small.

To calculate the \K -band extinction we use $R_V=3.1$ as recommeded by
SFD. In particular in the Galactic plane, though, $R_V$ can vary: 
towards the Galactic bulge it seems to be lower ($R_V=2.5$,
\citealt{nataf13}), while the centres of molecular clouds can show values
up to $4-6$. A change in $R_V$ results in a stretching of the \cer , but
only few data points would be affected (since we do not have many galaxies
in the Galactic Bulge area). 

%

\begin{table} 
\centering
\begin{minipage}{140mm}
\caption{{Comparison of $f$-values for different extinction laws 
for A7} \label{tabvar}}
\begin{tabular}{lccccc}
\hline
Colour & F99 &   C89 & Test & F99 & typical \\
       &     &       & & 2-digit precision & error \\
\hline
\jko  & 0.827 & 0.794 & 0.827 & 0.834 & 0.009 \\
\jho  & 0.817 & 0.871 & 0.838 & 0.817 & 0.006 \\
\hko  & 0.873 & 0.702 & 0.833 & 0.895 & 0.005 \\
\hline
\end{tabular}
\end{minipage}
\end{table}

\subsection{Correction for galaxy radius }

While smaller galaxies can be mostly or even fully contained in a
fixed-radius aperture, for the largest galaxies a small aperture may
contain only the bulge and therefore appear redder. Plotting colours versus
major axis radii we find a slight dependence of up to $+0.0007\pm0.0001$
for \jk -colours, with the worst case for PK and being generally larger for
the smaller apertures; ISO colours, on the other hand, show the reverse
effect with a slope of $-0.0004\pm0.0001$. The effect starts disappearing
when we remove all galaxies with radii smaller than around 15\arcsec.
Applying a correction for this dependence, however, has a negligible effect
on the \cer , mainly because the small galaxies are evenly distributed
across the range of extinctions. It should be noted that there is also a
dependence on morphological type on which we have little information. Ideally,
we should use an aperture size that includes only light from the bulges
(where available) and thus would vary from galaxy to galaxy. However, as
already mentioned, since the effect is negligible and other uncertainties
dominate the error budget, we have not applied a correction for this.


\subsection{Correction for stellar density }

In Paper I, we found that the colours show a slight dependence on stellar
densities. We have repeated the investigation for all apertures and find
that the effect on the determination of the \f\ is negligible.  While the
dependence of colour on star density is $-0.078\pm0.010$ for the isophotal
colours and smaller for the other apertures (except PK), \eg\ 
$-0.037\pm0.011$ for A7, the change in the \f\ is similar for all apertures
and smaller than the 1-$\sigma$ error of 0.009: for A7, $f$ increases from
$0.828$ to $f=0.835$ (for $N=3425$ since not all galaxies have a stellar
density defined). Since the effect is small and to avoid introducing an
additional uncertainty through the star density--colour relation, we have
not applied such a correction.

\subsection{k-correction }

Since redshift changes a galaxy's colour (in \jk\ they become redder), we
have applied a k-correction to those galaxies that have redshift
information and investigated the effect on the \f\ measurement. We use the
k-correction given by \citet{chilingarian10} who made their script
available online\footnote{http://kcor.sai.msu.ru/}. The k-correction
depends on the colour of a galaxy, but instead of using the individual
colours, we adopted generic colours: \jko\ $= 1\fm0$, \jho\ $= 0\fm7$,
\hko\ $= 0\fm3$ (see Paper I). This avoids additional uncertainties due to
errors in one of the passbands which slightly increase the scatter in the
\cer\ (from 0.122 to 0.128). The maximum difference in the two k-correction
versions is $\pm 0\fm02$ for \jk , with a mean difference of
$0\fm0003\pm0\fm0002$. The k-corrections and adopted velocities used are
given in a table described in the appendix.

\begin{figure} 
\centering
\includegraphics[width=0.47\textwidth]{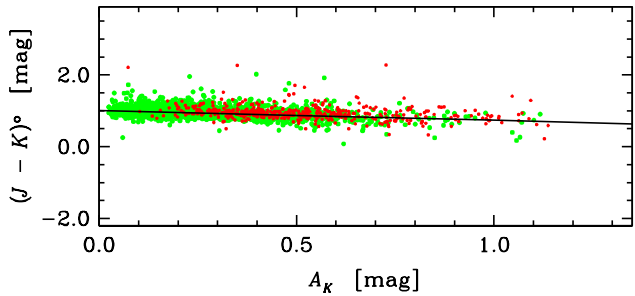}
\caption{Same as Fig.~\ref{colextplot}, where green dots depict galaxies
  with redshift information and red dots galaxies without redshifts. The
  regression line was fitted to the green dots.
}
\label{kcplot}
\end{figure}

We find that the k-correction has a negligible effect on the determination
of the \f . However, not all galaxies have a redshift measurement (about
13\% are still missing, see the red dots in Fig.~\ref{kcplot}), and the
reduced sample leads to a lower \f\ of $0.801\pm0.010$. The same trend is
found for all apertures. To understand better where the selection bias
comes from and how it may affect the result if we had redshift measurements
for all galaxies, we investigated further.

\begin{figure} 
\centering
\includegraphics[width=0.45\textwidth]{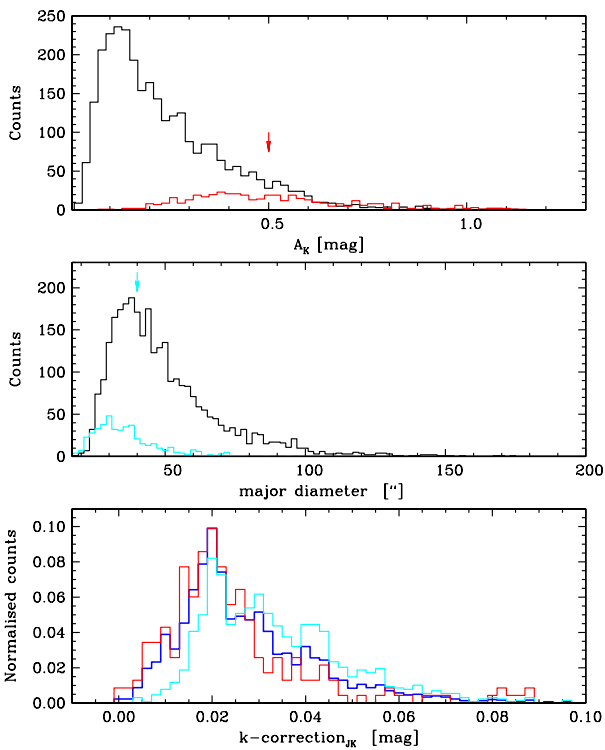}
\caption{Distributions of galaxies with extinction $A_{\rm K}$ (top), major
  axis diameter $a$ (in arcseconds, middle) and k-correction in \jk
  (bottom). The full sample is shown in black, galaxies with redshift
  information in dark blue, and galaxies with no redshifts in red (top) and
  cyan (middle), respectively. The bottom panel shows the samples after the
  application of the respective cuts indicated by the arrows and colours in
  the top and middle panel. For more information see text.
 }
\label{kchistplot}
\end{figure}

First, we looked at the distribution of galaxies with and without redshift
measurements as a function of extinction and major axis diameter, both of
which have a likely influence on whether a redshift was attempted to be
measured. The top two panels in Fig.~\ref{kchistplot} show the respective
distributions for the full sample in black and for the sample without
redshift information in colour. As expected, redshifts are predominantly
lacking for galaxies at higher extinctions and having smaller observed
diameters. The bottom panel, showing normalised counts for better
comparison, presents the distribution of the k-correction values for all
galaxies with a redshift (in dark blue) as well as for those after the
application of an additional cut in extinction (red: \ak $>0\fm5$) or in
diameter (cyan: $a<40\arcsec$) applied, as indicated by the arrows in the
upper panels. While the blue and red histograms compare reasonably well,
the cyan histogram shows a clear shift (by about $0\fm005$) with respect to
the blue one. This means that galaxies at high extinctions do not show any
preference in distance (as expected) and, consequently, in k-correction,
but smaller galaxies tend to have a slightly higher k-correction because
many are more distant, normal-sized galaxies. Nonetheless, the smaller
galaxies are evenly distributed over extinctions and therefore do not
introduce an extinction-dependent bias in the \cer .


%

\begin{figure} 
\centering
\includegraphics[width=0.40\textwidth]{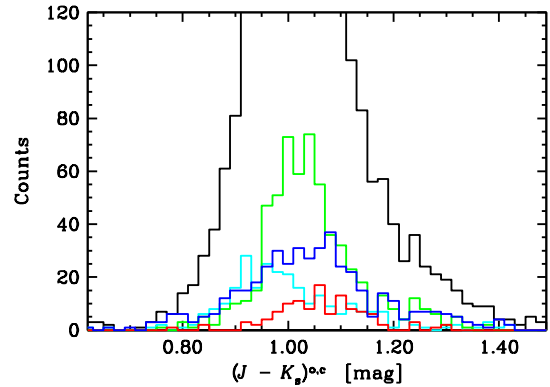}
\caption{Colour distribution of galaxies with various velocity ranges: all
  galaxies (black), $cz < 3000$\kms\ (cyan), 8000\kms\ $ < cz <$
  11000\kms\ (green), 15000\kms\ $ < cz <$ 25000\kms\ (red), no velocity
  (dark blue).
 }
\label{histcolplot}
\end{figure}

Secondly, we can study the colour distribution of these samples.
Figure~\ref{histcolplot} shows the lower part of the histogram in colour
\jkoc , where we applied $f=0.83$ to the extinction correction to be as
close to the intrinsic colour as possible. The figure shows various
redshift ranges in colour (low to high redshifts going from cyan through
green to red, with the full sample depicted in black), together with the
galaxies that have no velocity information in dark blue. The median \jkoc
-values go from $0\fm981$ through $1\fm034$ to $1\fm054$, respectively,
while the no-redshift sample has a median of $1\fm044$, albeit with a
larger scatter of $0\fm20$ (as compared to $0\fm10-0\fm16$ for the other
samples). Though NIR colours are only a rough indicator of redshift (\eg
\citealt{jarrett04}), this shows that the missing galaxies seem to be well
distributed across redshift space with no likely bias in distance (or
k-correction).

We thus conclude that though the sub-sample of galaxies with redshifts
gives a different \f , the k-correction itself has a negligible effect on
its determination: (1) the missing galaxies seem to have a similar redshift
distribution to those with redshifts, with a moderate lack of very close-by
galaxies ($cz \la 3000$\kms ), (2) the amount of k-correction does not show
any correlation with extinction for this sample, and (3) the missing
galaxies come mainly from the intermediate-to-high extinction range which
severely depletes the \cer\ in these areas, likely explaining the deviant
\f . In fact, we ought to adjust the upper limit in \ak\ (as discussed in
Sec.~\ref{ablimit}) to about \ak $< 1\fm04$ (at the end of the new
`plateau'), leading to a slight increase of the \f\ to $0.809\pm 0.010$,
which deviates by only 2$\sigma$ from the original value.

\subsection{Bootstrapping}

In Sec.~\ref{outliers}, we investigated the effect of the remaining
outliers on the linear regression and found it to be negligible. To look
further into the effects of sample selection on the result, we applied the
bootstrap method, excluding a third of randomly selected data points. We
ran this 20 times and found a mean $f=0.831 \pm 0.002$ with individual
values ranging from 0.818 to 0.848. This agrees well with the \f\ for our
cleaned sample, $f=0.827\pm0.009$ and we conclude that our result is robust
with respect to sample selection and outliers.

\subsection{Discussion}

In conclusion we can say that our derivation of a correction factor to the
SFD maps, $f=0.83\pm0.01$, is very robust. The largest variation is found
across the colours (see Table~\ref{tabvar}), which seem to be sensitive to
the adopted extinction law. The choice of aperture size has a small effect,
with \f s ranging from 0.823 (S5) to 0.832 (A5), see Table~\ref{tabfvalue}.
Additional sample selections (as discussed in Secs~\ref{2mzoa}
and~\ref{outliers}) have a similar impact ($0.815 - 0.836$ across all
apertures, see minimum and maximum values in Table~\ref{tabfvalue}). A7 and
A5 seem to be the most robust, while ISO and PK seem to be more vulnerable
(they also have the least objects in the sample). It is notable that even
the isophotal colours, much more subject to insufficient star subtraction,
show little deviation and agree well with our final value of
$f=0.827\pm0.009$. The effects of corrections for radii, stellar densities
and k-correction are negligible. In the latter case, the galaxies without
redshift (13\% of the cleaned sample) cause a bias since mostly
higher extinction objects are lacking, resulting in a lower \f\ of
$0.801\pm0.010$. The bias is expected to disappear once all galaxies have
a redshift measured. The nominal uncertainty on our \f\ is 0.009, and,
based on the variations found, we do not expect the final error to be
higher than 0.01.

\begin{table} 
\centering
\begin{minipage}{140mm}
\caption{{Comparison of parameters for different apertures} \label{tabfvalue}}
\begin{tabular}{lcccccc}
\hline
 & A5 & A7 & A10 & ISO & S5 & PK \\
\hline
\f\           & 0.832 & 0.827 & 0.829 & 0.826 & 0.823 & 0.828 \\
$f_{\rm min}$ & 0.825 & 0.822 & 0.824 & 0.821 & 0.815 & 0.823 \\
$f_{\rm max}$ & 0.832 & 0.830 & 0.836 & 0.835 & 0.827 & 0.831 \\
Scatter       &$0\fm141$&$0\fm134$&$0\fm130$&$0\fm129$&$0\fm134$&$0\fm173$\\
$N$           &  \phantom{.}3447 &  \phantom{.}3460 &  \phantom{.}3395 &  \phantom{.}3272 &  \phantom{.}3391 &  \phantom{.}3391 \\ 
%
\hline
\multicolumn{7}{p{8cm}}{Note: All values have an error of 0.009 except for  
 PK which has an error of 0.012.} 
\end{tabular}
\end{minipage}
\end{table}

\section{Other extinction maps }  \label{planck}

The analysis we have shown so far was based on the all-sky extinction map
derived from DIRBE/IRAS data by SFD which is the most widely used to date
(recently with the SF11 modulation). Recently, various extinction maps
based on Planck satellite data have become available. We can now use our
cleaned sample to compare their quality and whether they may also need a
correction factor. In particular, we investigated the following four maps,
all of which have a spatial resolution of $6\farcm1$ (\ie comparable to
SFD):
\begin{description}

\item {\bf P-PR1} The first Planck extinction map \citep{planck14} is based
  on the PR1 data from 2013 and the IRAS $100\mu$m data. A modified
  blackbody model was used to fit the data.

\item {\bf P-MF} \citet{meisner15} also used the Planck PR1 and DIRBE/IRAS
  $100\mu$m data. Instead of a modified blackbody model, they used a
  two-component model. Their map is given in optical depth at a frequency
  of 545\,GHz ($\lambda=550\mu$m), with a provided conversion of \ebv $=
  2624.4472 \cdot {\rm TAU}545 - 0.00260618$.

\item {\bf P-AV} \citeauthor{planck16a} (2016a) fitted a physical dust
  model from \citet{dl07} to the full mission Planck PR2 (2015) data, WISE
  $12\mu$m data and IRAS $60\mu$m as well as $100\mu$m data. Their
  extinction map is given in \av\ (their renormalised AV\_RQ), which we
  converted to \ebv\ using $R_{\rm V} = 3.1$.


\item {\bf P-GNILC} \citeauthor{planck16b} (2016b) used the PR2 Planck data
  in combination with the IRAS $100\mu$m map and took special care to
  reduce the cosmic infrared background contamination in the data, using
  the generalised needlet internal linear combination (GNILC)
  component-separation method. Their map is given in dust optical depth at
  353\,GHz ($\lambda=849\mu$m), with a provided conversion of \ebv $= 14900
  \cdot {\rm TAU}353$.


\end{description}
We have not used other available maps because they either have lower
spatial resolutions, do not cover the whole Galactic plane (\eg
\citealt{nidever12}), or are 3-dimensional, that is, do not cover the full
line-of-sight out of our Galaxy (\eg \citealt{gonzalez12}).


\begin{figure} 
\centering
\includegraphics[width=0.47\textwidth]{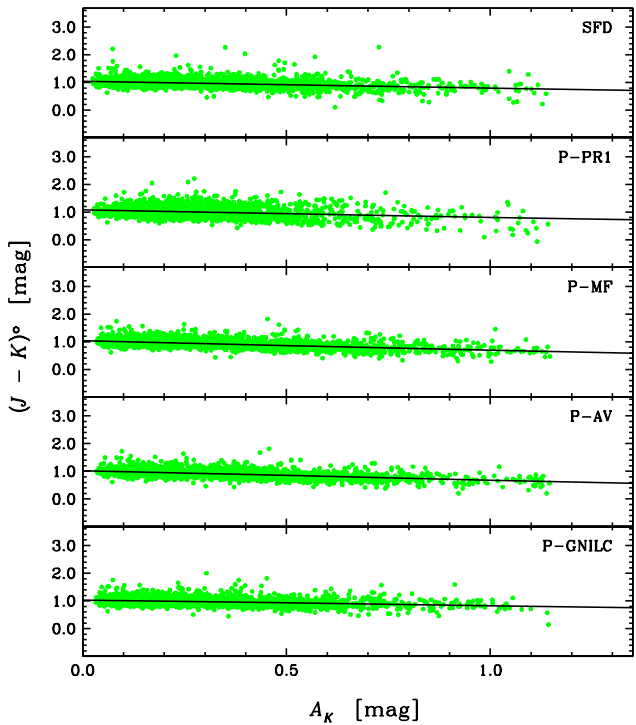}
\caption{Same as Fig.~\ref{colextplot} for various extinction maps as
  indicated by the labels.
}
\label{colextplplot}
\end{figure}

To compare the Planck maps with the SFD map, we have used the
\cer\ (Fig.~\ref{colextplplot}), assuming the scatter to be an indicator of
the quality of the extinction values since all other parameters are kept
unchanged. As mentioned above, the SFD \cer\ has a scatter of $0\fm134$. In
comparison, the P-PR1 map is the worst with a considerably larger scatter
of $0\fm175$. The other three maps lead to a similarly reduced scatter
between $0\fm116$ and $0\fm119$ (see Table~\ref{compmapstab}), indicating
that both improved methods and improved data are of advantage. This
comparison is also valid if we consider only galaxies with $|b| \ge 5\degr$
where the SFD maps are fully calibrated.

Interestingly, all relations show still a considerable dependence on
extinction, implying the continued need of a correction factor similar to
that for the SFD map: the P-GNILC map shows the least dependence on
extinction with $f=0.859\pm 0.008$, while both P-AV and P-MF show the
highest dependence ($f=0.759\pm 0.007$ and $f=0.760\pm 0.007$,
respectively). The offset (that is, the \jk -colour at zero extinction) is
comparable for all maps except for P-PR1, which is significantly redder at
$1\fm080\pm 0\fm005$.

We conclude that the three maps P-MF, P-AV and P-GNILC are preferable over
the SFD map, while the P-PR1 map appears unreliable.

\begin{table} 
\centering
\begin{minipage}{140mm}
\caption{{Comparison of parameters for different extinction maps} \label{compmapstab}}
\begin{tabular}{lccc}
\hline
Extinction map & \f & scatter & offset \\
               &    & [mag]   & [mag]  \\
\hline
SFD     & $0.827 \pm 0.009$ & 0.134 & $1.036 \pm 0.004$ \\
P-PR1   & $0.808 \pm 0.013$ & 0.175 & $1.080 \pm 0.005$ \\
P-MF    & $0.760 \pm 0.007$ & 0.118 & $1.031 \pm 0.004$ \\
P-AV    & $0.759 \pm 0.007$ & 0.116 & $1.017 \pm 0.00$4 \\
P-GNILC & $0.859 \pm 0.008$ & 0.119 & $1.028 \pm 0.004$ \\
\hline
\end{tabular}
\end{minipage}
\end{table}


\section{Variation across the ZoA}  \label{maps}

\begin{figure*} 
\centering
\includegraphics[width=0.97\textwidth]{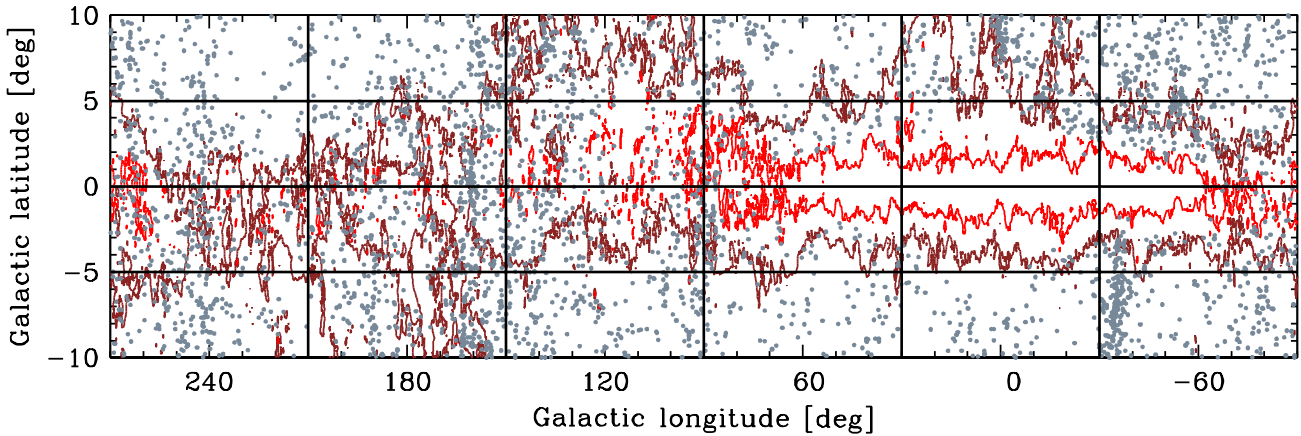}
\caption{Sky map with contours of Galactic extinction ($A_{\rm K}=0\fm3$ in
  brown and $A_{\rm K}=1\fm0$ in red; from SFD). Galaxies are shown as grey
  dots. The grid depicts the individual cells used in the
  analysis. 
}
\label{lbmapplot}
\end{figure*}

SFD give the caveat that their extinction maps are not calibrated at
\mbox{$|b|<5\degr$}, partly because of low galaxy counts, partly
because of unsubtracted FIR sources (see discussion in Paper I), and
partly because of badly handled artefacts in the temperature map. Our
sample is large enough that we can divide the ZoA into smaller cells to
look for any large-scale variation with both Galactic longitude and
latitude.  Figure~\ref{lbmapplot} shows the distribution of the sample
galaxies in Galactic coordinates (grey dots) with contours of Galactic
extinction overlaid as indicated in the caption. The individual cells
($60\degr \times 5\degr$) are also shown. It is obvious that some cells
have only few galaxies (around the Galactic bulge), while other cells show
only little variation in extinction.

\begin{figure*} 
\centering
\includegraphics[width=0.7\textwidth]{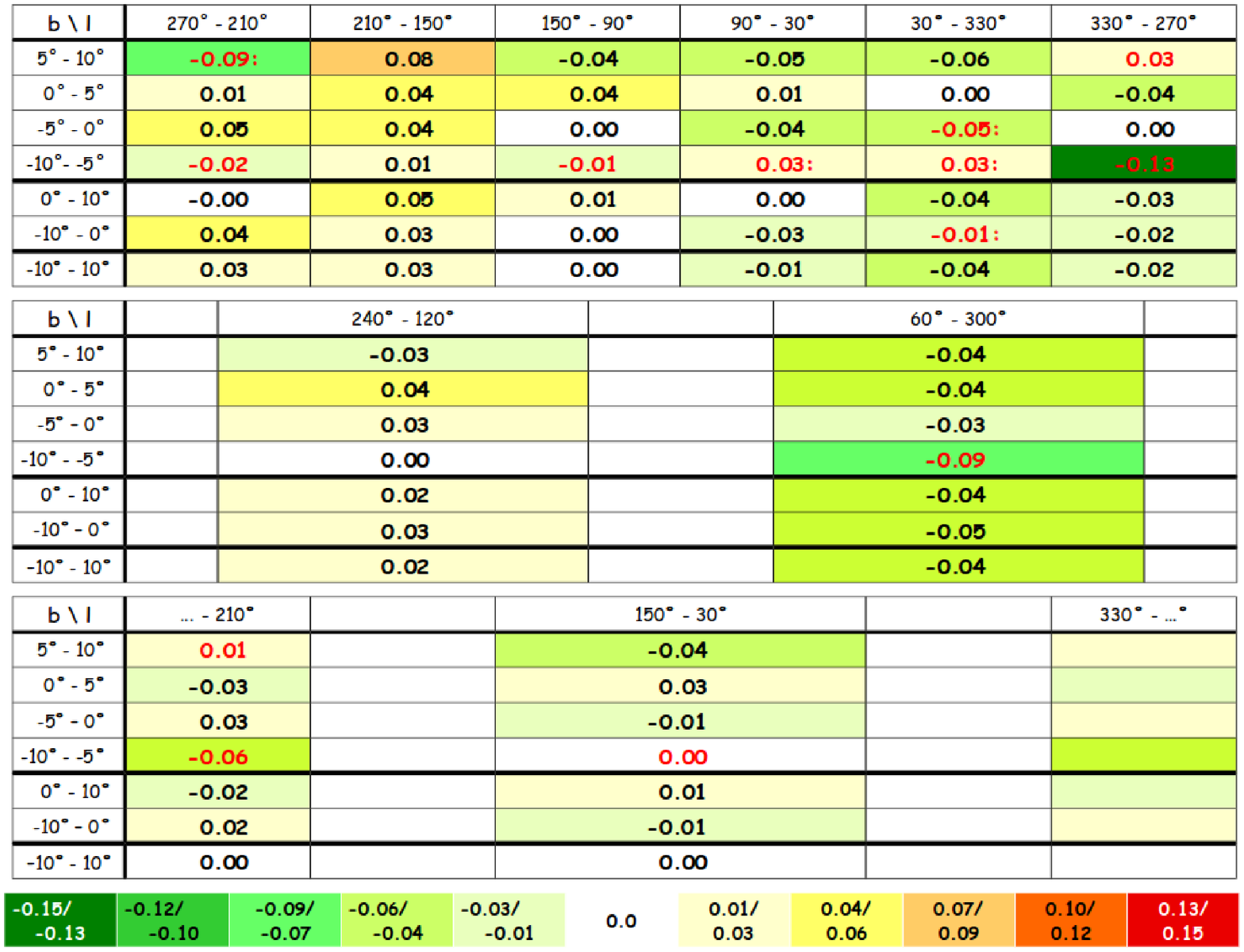}
\caption{Binned sky map with differential \f s ($f_{\rm cell} - f_{\rm
    full}$) per cell. The colour scale is shown at the bottom. Red numbers
  may be overestimated and colons indicate uncertain values. The $330\degr
  - 210\degr $ cells are wrapped around the Galactic Centre.
}
\label{lbmap2plot}
\end{figure*}

To minimise the uncertainties in the slope of the \cer\ for the individual
cells, in particular where the extinction range is small, we kept the
intercept (\ie the unobscured colour) fixed at the value derived for the
full sample. We then determined the \f\ for each cell, keeping an eye on
errors, scatter and extinction range and using the bootstrap method to
further estimate the reliability. To avoid the above mentioned problems of
the small cell size, we also binned up cells in various ways to search for
reliable patterns.

We find that the \f\ varies across the ZoA between $0.70\pm0.04$ and
$0.91\pm0.03$ or, in more useful terms, the deviation $\Delta f$ from the
full sample \f\ ($f_{\rm cell} - f_{\rm full}$) goes from $-0.13$ to
$0.08$. We have visualised the results in Fig.~\ref{lbmap2plot} by using a
colour map of the cells from dark green (most negative) to red (most
positive). Numbers in red denote a short extinction range (with a maximum
of \ak$=0\fm55$) to indicate that the \f\ is possibly overestimated (see
Fig~\ref{ablimplot}). Values with a colon are deemed less reliable due to
larger errors, small number counts or large variation when using the
bootstrap method. Despite these problems there appears to be a pattern,
mainly along Galactic longitudes. Hence, we have binned the cells to
emphasise the Galactic Centre region ($-60\degr < l < +60\degr$) versus the
Anticentre region ($120\degr < l < 240\degr$) as shown in the lower panels
of the figure. It is now apparent that the \f\ towards the Galactic Centre
is somewhat lower than towards the Anticentre region (by about 0.06 or
6$\sigma$ when binning over all latitudes). A possible variation with
latitude is less clear and subject to higher uncertainties in the lowest
latitude bin ($-10\degr < b < -5\degr$). We also looked for a possible
variation with the Cosmic Microwave Background (CMB) dipole and used
$120\degr$-wide latitude bins around $l=270\degr$ and $l=90\degr$. The
differences here are negligible and reveal no clear pattern.

\subsection{Comparative analysis of the available extinction maps}

\begin{figure} 
\centering
\includegraphics[width=0.40\textwidth]{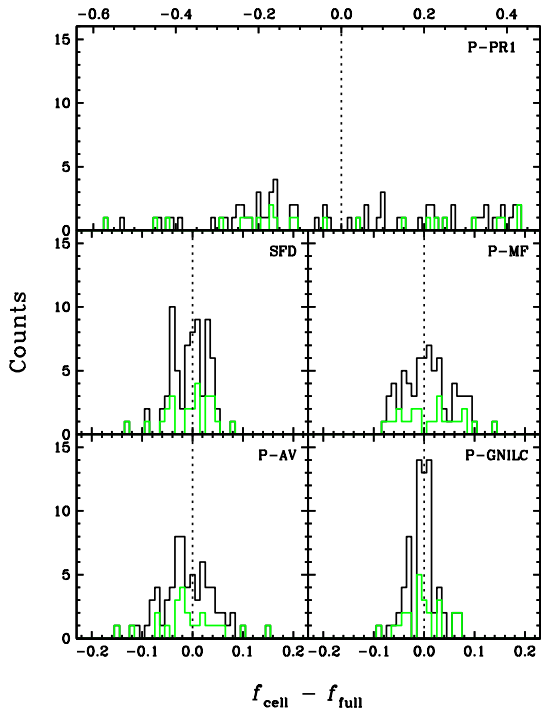}
\caption{Histograms of $\Delta f$ for all extinction maps, as
  labelled. Green histograms indicate the individual cells ($60\degr \times
  5\degr$) and black histograms comprise all cells analysed. Dotted
  vertical lines indicate $\Delta f = 0$.
}
\label{histdfplot}
\end{figure}

\begin{figure} 
\centering
\includegraphics[width=0.35\textwidth]{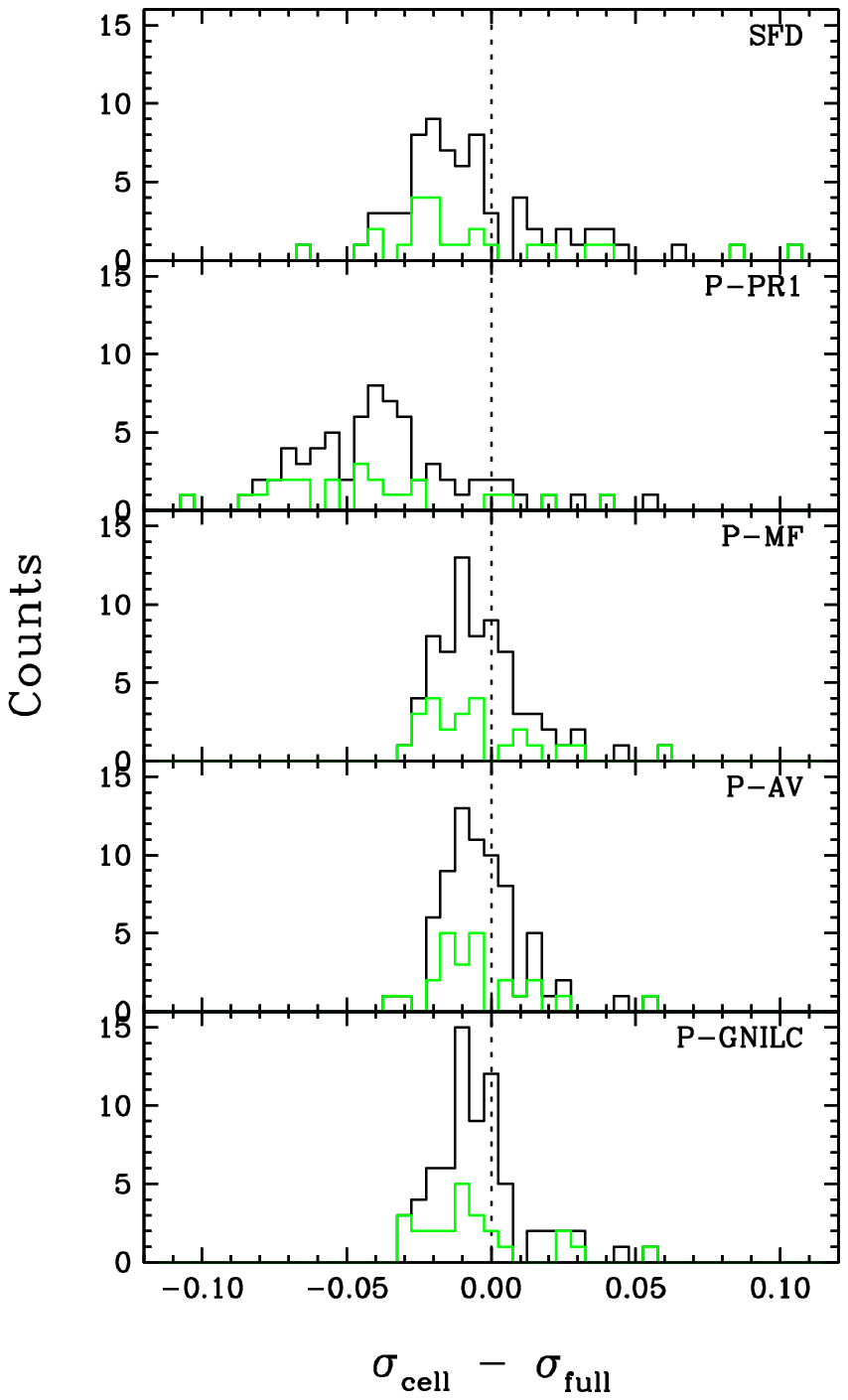}
\caption{Same as Fig.~\ref{histdfplot} for the differential scatter in the
  \cer .
}
\label{histdsplot}
\end{figure}

We have applied the same analysis to the other extinction maps to see
whether they also reveal a variation with position in the sky. For easier
comparison, Figs~\ref{histdfplot} and~\ref{histdsplot} show histograms of
the differences in \f s and scatter (green histograms denote the 24 small,
independent cells, black stands for all 70 cells used). The individual
maps, in the style of Fig.~\ref{lbmap2plot}, are given in the Appendix.

In the case of P-PR1 we find much larger variations of the \f\ with more
extreme values, varying from 0.24 to 1.24 ($-0.57 \le \Delta f \le
0.43$). The scatter of the individual \cer s, on the other hand, is usually
much smaller than for the full sample, which, together with the larger
variation in slopes, explains the higher scatter of the full relation.
Interestingly, the emerging pattern is similar to the SFD map, with the
Galactic Centre region having lower \f s than the Anticentre region, in
this case by the considerable amount of 0.59. As to variations with
latitude, there seems to be a small decrease in the \f\ in the Centre
region from above to below the plane by about 0.18. Considering the large
variation from cell to cell, though, this may not be indicative of a
coherent pattern, in particular since the other longitude cells do not show
a similar change. The cells depicting the CMB dipole also show a small
difference ($\Delta f = 0.17$), but considering the map of the individual
cells this seems only to reflect the much stronger Centre--Anticentre
juxtaposition.

The analysis of the improved Planck PR1 map P-MF shows similar variations
as SFD and reflects the small scatter of the full sample \cer .  The \f s
vary from 0.68 to 0.90 ($-0.08 \le \Delta f \le 0.14$). Again, there is a
dichotomy between the Galactic Centre and Anticentre regions, but, contrary
to the SFD maps, the Centre region shows {\it higher} values than the
Anticentre region, by about 0.05 or 5$\sigma$. As with the SFD maps, there
is no clear variation with latitude or the CMB dipole.

The P-AV map, based on Planck PR2 data, shows slightly larger variations
than SFD, going from 0.61 to 0.91 ($-0.15 \le \Delta f \le 0.15$). As for
SFD, the Galactic Centre region shows lower values than the Anticentre
region, but more pronounced with a difference of 0.08 or 8$\sigma$. As for
P-PR1, there appears to be a possible decrease in the \f\ with latitude in
the Galactic Centre region (by 0.05 or 5$\sigma$), but the same caveat
applies here: variations from cell to cell are of similar order (albeit
with larger errors), and the Anticentre region does not show any
pattern. The difference with respect to the CMB dipole only reflects the
stronger Centre--Anticentre juxtaposition.

The P-GNILC map shows the least variations, going from 0.77 to 0.93 ($-0.09
\le \Delta f \le 0.07$). In this case, no pattern can be discerned. The
difference between the Galactic Centre and Anticentre region is 0.01
(1$\sigma$) and 0.03 (3$\sigma$) for the CMB dipole. No latitude dependence
could be discerned either.

Considering Figs~\ref{histdfplot} and~\ref{histdsplot}, P-PR1 shows
strikingly different behaviour compared to the other maps: there is no
preferred value for $f$ at all, instead it varies evenly across the whole
spread of values. The values of the scatter of the individual cells, on the
other hand, are markedly smaller, confirming that the large scatter of the
full sample is either due to varying background contamination or to
insufficient calibration anchors across the ZoA. In fact, the other four
maps also show a slightly smaller scatter for the individual cells, which
likely means that the variation in the \f\ is real. The $\Delta f$ of these
four maps show the more expected grouping around zero, though only the
P-GNILC map shows a pronounced peak at zero for both the individual cells
as well as for the binned ones. P-AV, in particular, seems to show a
bipolar distribution instead.

It is worth noting that for the map with the least variations and smallest
correction factor, P-GNILC, particular care was given by the authors to the
removal of any residuals due to the anisotropy of the CMB. Based on the
overall comparison, we recommend the use of this map in preference to the
SFD map.

\section{Summary and Conclusions}  \label{concl}

In this paper, we have used 2MASS NIR colours of galaxies to calibrate the
DIRBE/IRAS extinction map by SFD at low Galactic latitudes. We have derived
a correction factor $f = 0.83\pm0.01$ by fitting a linear regression to the
\cer , using 3460 galaxies.

In the first part of the paper, we concentrated on finding the best
parameter settings and corrections for a clean analysis, using the widely
used SFD DIRBE/IRAS maps. We found that the \jk -colour from the
$7\arcsec$-radius aperture is most useful, balancing the lower photometric
uncertainty of a small aperture, little affected by contamination with
stars, against the need to retain an as large as possible sample for the
subsequent analysis. We therefore applied only the most necessary cuts in
photometric errors and flag settings.

One of the problems of a reliable \cer\ is the coarse spatial resolution of
the extinction maps which introduces a bias in the {\it finding} of
galaxies, that is, galaxies are predominantly found in lower extinction
pockets while the average extinction of that area is higher. As a
consequence, galaxy colours will appear systematically too blue since the
extinction correction applied is too high. We found this to be mainly a
problem in high extinction areas where the extinction often varies rapidly
due to dark clouds. This is confirmed in the \cer\ that shows the galaxy
colours deviating from the linear relation at the higher extinction end. We
have thus adopted an upper extinction limit of \ak$=1\fm15$ (or \ebv$\simeq
3\fm19$).

We investigated the effects of using different extinction laws by comparing
the \f\ derived for the three colours \jk , \jh\ and \hk . We have found
that the F99 extinction law shows better agreement than the law derived by
C89 but note that the coefficient used for the \HH -band could still be
improved.

We tested the application of the k-correction to a sub-sample of the
galaxies with redshift information. While the k-correction itself does not
have any effect on the \f , the subsample itself leads to a lower \f\ of
$0.80\pm0.01$.  This is mainly due to a selection bias in redshift
measurements such that smaller galaxies at higher extinctions are less
likely to have a redshift measurement. We note, though, that since our
sample is magnitude limited with \Ko\,$ \le 11\fm25$, it comprises
predominantly near-by galaxies and the spread in k-correction for our
sample is small.

The derived correction factor $f=0.83\pm0.01$ is slightly smaller than the
one found by SF11 ($f=0.86$) for the SFD maps. Since we agree with SF11 on
applying the F99 reddening law and using $R_V = 3.1$, the 3-$\sigma$
difference (if real) must have other reasons. We have shown that the
restriction to a small extinction range (\ak$<0\fm4$) can lead to slightly
higher values with larger uncertainties. SF11 note that they have only
little data beyond \ak$=0\fm11$. We thus recommend that at low latitudes
($|b| < 10\degr$) or high extinctions the application of $f=0.83$ is used
instead.



We used our sample to test four other all-sky extinction maps, all based on
Planck data. We did not investigate other available maps since they either
have a lower spatial resolution, cover only a limited region in the sky or
measure extinctions towards stars, which underestimates the full
line-of-sight extinction to the extragalactic sky. Next to the original
Planck extinction map, based on PR1 data \citep{planck14}, we used a
modified map of the PR1 data by \citet{meisner15}. Planck PR2 data were
used by \citeauthor{planck16a} (2016a) and \citeauthor{planck16b}
(2016b). We found that all maps require a correction factor similar to the
one for the SFD maps. Furthermore, the original Planck map (P-PR1) shows a
larger scatter and a very high offset in the relation, that is, an
unobscured \jk\ colour that is too red.

Our subsequent analysis of a possible variation of the correction factor
over Galactic longitude and latitude revealed more problems with the P-PR1
map where the correction factor varies strongly, seemingly even within some
of the $60\degr \times 5\degr$ cells. Furthermore we found that all but the
P-GNILC map show a small but marked difference between the Galactic Centre
and Anticentre regions, and none with the direction of the CMB dipole.

We conclude by recommending the use of the P-GNILC map at low latitudes as
showing the least variation across the sky and requiring the smallest
correction, $f=0.86\pm0.01$.

While our NIR ZoA galaxy sample is the largest and most complete to date,
it still suffers from missing galaxies in the highest stellar density
region, \ie the Galactic bulge. With the new NIR large-area surveys
UKIDSS-GPS \citep{lucas08} and VVV (\eg\ \citealt{minniti10}) we can fill
in the gap and increase the sample size, which will allow a finer binning
and thus a more smoothly varying correction factor with longitude.

We also plan to compare in detail the full-sky maps with regional 3D maps
and to investigate ways to calibrate the 3D maps to the full line-of-sight
extinctions displayed in the coarser maps while retaining the higher
spatial resolution information. This may only be possible, though, where
stars at the far side of the Galactic Bulge were used.

%

\section*{Acknowledgments}

The authors thank the observing teams in our project for their efforts and
for making spectroscopic data available prior to publications: L.\ Macri,
T.\ Lambert, K.\ Said. This publication makes use of data products from the
Two Micron All Sky Survey, which is a joint project of the University of
Massachusetts and the Infrared Processing and Analysis Center, funded by
the National Aeronautics and Space Administration and the National Science
Foundation. This research also has made use of the HyperLeda database, the
SIMBAD database, operated at CDS, Strasbourg, France, the NASA/IPAC
Extragalactic Database (NED) which is operated by the Jet Propulsion
Laboratory, California Institute of Technology, under contract with the
National Aeronautics and Space Administration and the Sloan Digital Sky
Survey which is managed by the Astrophysical Research Consortium for the
Participating Institutions. RKK's research is supported by the South
African Research Chair Initiative of the Department of Science \&
Innovation and the National Research Foundation (NRF).  ACS thanks the
South African NRF for their financial support.

\section*{Data Availability}
The data underlying this article are available in the article and in its
online supplementary material.

\bibliographystyle{mn2e} 
\bibliography{ZoA_bibfile_long} 

\begin{thebibliography}{}

\bibitem[\protect\citeauthoryear{{Acker}, {Stenholm} \& {Veron}}{{Acker}
  et~al.}{1991}]{1991AuAS...87..499A}
{Acker} A.,  {Stenholm} B.,    {Veron} P.,  1991, \aaps, 87, 499

\bibitem[\protect\citeauthoryear{{Alam}, {Albareti}, {Allende Prieto},
  {Anders}, {Anderson}, {Anderton}, {Andrews}, {Armengaud}, {Aubourg}
  et~al.,}{{Alam} et~al.}{2015}]{2015ApJS..219...12A}
{Alam} S.,  {Albareti} F.~D.,  {Allende Prieto} C.,  {Anders} F.,  {Anderson}
  S.~F.,  {Anderton} T.,  {Andrews} B.~H.,  {Armengaud} E.,  {Aubourg} {\'E}.,
    et~al., 2015, \apjs, 219, 12

\bibitem[\protect\citeauthoryear{{Ali}, {Dopita}, {Basurah}, {Amer}, {Alsulami}
  \& {Alruhaili}}{{Ali} et~al.}{2016}]{2016MNRAS.462.1393A}
{Ali} A.,  {Dopita} M.~A.,  {Basurah} H.~M.,  {Amer} M.~A.,  {Alsulami} R.,
  {Alruhaili} A.,  2016, \mnras, 462, 1393

\bibitem[\protect\citeauthoryear{{Beaulieu}, {Dopita} \& {Freeman}}{{Beaulieu}
  et~al.}{1999}]{1999ApJ...515..610B}
{Beaulieu} S.~F.,  {Dopita} M.~A.,    {Freeman} K.~C.,  1999, \apj, 515, 610

\bibitem[\protect\citeauthoryear{{Beuing}, {Bender}, {Mendes de Oliveira},
  {Thomas} \& {Maraston}}{{Beuing} et~al.}{2002}]{2002AuA...395..431B}
{Beuing} J.,  {Bender} R.,  {Mendes de Oliveira} C.,  {Thomas} D.,
  {Maraston} C.,  2002, \aap, 395, 431

\bibitem[\protect\citeauthoryear{{Bicay} \& {Giovanelli}}{{Bicay} \&
  {Giovanelli}}{1986a}]{1986AJ.....91..705B}
{Bicay} M.~D.,  {Giovanelli} R.,  1986a, \aj, 91, 705

\bibitem[\protect\citeauthoryear{{Bicay} \& {Giovanelli}}{{Bicay} \&
  {Giovanelli}}{1986b}]{1986AJ.....91..732B}
{Bicay} M.~D.,  {Giovanelli} R.,  1986b, \aj, 91, 732

\bibitem[\protect\citeauthoryear{{Bicay} \& {Giovanelli}}{{Bicay} \&
  {Giovanelli}}{1987}]{1987AJ.....93.1326B}
{Bicay} M.~D.,  {Giovanelli} R.,  1987, \aj, 93, 1326

\bibitem[\protect\citeauthoryear{{Bikmaev}, {Burenin}, {Revnivtsev}, {Sazonov},
  {Sunyaev}, {Pavlinsky} \& {Sakhibullin}}{{Bikmaev}
  et~al.}{2008}]{2008AstL...34..653B}
{Bikmaev} I.~F.,  {Burenin} R.~A.,  {Revnivtsev} M.~G.,  {Sazonov} S.~Y.,
  {Sunyaev} R.~A.,  {Pavlinsky} M.~N.,    {Sakhibullin} N.~A.,  2008, Astronomy
  Letters, 34, 653

\bibitem[\protect\citeauthoryear{{Bottinelli}, {Durand}, {Fouque}, {Garnier},
  {Gouguenheim}, {Paturel} \& {Teerikorpi}}{{Bottinelli}
  et~al.}{1992}]{1992AuAS...93..173B}
{Bottinelli} L.,  {Durand} N.,  {Fouque} P.,  {Garnier} R.,  {Gouguenheim} L.,
  {Paturel} G.,    {Teerikorpi} P.,  1992, \aaps, 93, 173

\bibitem[\protect\citeauthoryear{{Bottinelli}, {Durand}, {Fouque}, {Garnier},
  {Gouguenheiml.}, {Loulergue}, {Paturel}, {Petit} \&
  {Teerikorpi}}{{Bottinelli} et~al.}{1993}]{1993AuAS..102...57B}
{Bottinelli} L.,  {Durand} N.,  {Fouque} P.,  {Garnier} R.,  {Gouguenheiml.}
  {Loulergue} M.,  {Paturel} G.,  {Petit} C.,    {Teerikorpi} P.,  1993, \aaps,
  102, 57

\bibitem[\protect\citeauthoryear{{Bottinelli}, {Gouguenheim}, {Loulergue},
  {Martin}, {Theureau} \& {Paturel}}{{Bottinelli}
  et~al.}{1994}]{1994ASPC...67..225B}
{Bottinelli} L.,  {Gouguenheim} L.,  {Loulergue} M.,  {Martin} J.~M.,
  {Theureau} G.,    {Paturel} G.,  1994, in {Balkowski} C.,  {Kraan-Korteweg}
  R.~C.,  eds, Unveiling Large-Scale Structures Behind the Milky Way Vol.~67 of
  Astronomical Society of the Pacific Conference Series, {An HI-Search for IRAS
  Galaxies in the Galactic Plane}.
p.~225

\bibitem[\protect\citeauthoryear{{Bottinelli}, {Gouguenheim} \&
  {Paturel}}{{Bottinelli} et~al.}{1981}]{1981AuAS...44..217B}
{Bottinelli} L.,  {Gouguenheim} L.,    {Paturel} G.,  1981, \aaps, 44, 217

\bibitem[\protect\citeauthoryear{{Bronfman}, {Nyman} \& {May}}{{Bronfman}
  et~al.}{1996}]{1996AuAS..115...81B}
{Bronfman} L.,  {Nyman} L.~A.,    {May} J.,  1996, \aaps, 115, 81

\bibitem[\protect\citeauthoryear{{Burbidge} \& {Burbidge}}{{Burbidge} \&
  {Burbidge}}{1972}]{1972ApJ...172...37B}
{Burbidge} E.~M.,  {Burbidge} G.~R.,  1972, \apj, 172, 37

\bibitem[\protect\citeauthoryear{{Burenin}, {Mescheryakov}, {Revnivtsev},
  {Sazonov}, {Bikmaev}, {Pavlinsky} \& {Sunyaev}}{{Burenin}
  et~al.}{2008}]{2008AstL...34..367B}
{Burenin} R.~A.,  {Mescheryakov} A.~V.,  {Revnivtsev} M.~G.,  {Sazonov} S.~Y.,
  {Bikmaev} I.~F.,  {Pavlinsky} M.~N.,    {Sunyaev} R.~A.,  2008, Astronomy
  Letters, 34, 367

\bibitem[\protect\citeauthoryear{{Burstein}, {Davies}, {Dressler}, {Faber},
  {Stone}, {Lynden-Bell}, {Terlevich} \& {Wegner}}{{Burstein}
  et~al.}{1987}]{burstein87}
{Burstein} D.,  {Davies} R.~L.,  {Dressler} A.,  {Faber} S.~M.,  {Stone} R.
  P.~S.,  {Lynden-Bell} D.,  {Terlevich} R.~J.,    {Wegner} G.,  1987, \apjs,
  64, 601

\bibitem[\protect\citeauthoryear{{Burstein} \& {Heiles}}{{Burstein} \&
  {Heiles}}{1978}]{burstein78}
{Burstein} D.,  {Heiles} C.,  1978, ApJ, 225, 40

\bibitem[\protect\citeauthoryear{{Burstein} \& {Heiles}}{{Burstein} \&
  {Heiles}}{1982}]{burstein82}
{Burstein} D.,  {Heiles} C.,  1982, AJ, 87, 1165

\bibitem[\protect\citeauthoryear{{Cardelli}, {Clayton} \& {Mathis}}{{Cardelli}
  et~al.}{1989}]{cardelli89}
{Cardelli} J.~A.,  {Clayton} G.~C.,    {Mathis} J.~S.,  1989, ApJ, 345, 245

\bibitem[\protect\citeauthoryear{{Chamaraux}, {Cayatte}, {Balkowski} \&
  {Fontanelli}}{{Chamaraux} et~al.}{1990}]{1990AuA...229..340C}
{Chamaraux} P.,  {Cayatte} V.,  {Balkowski} C.,    {Fontanelli} P.,  1990,
  \aap, 229, 340

\bibitem[\protect\citeauthoryear{{Chamaraux}, {Masnou}, {Kaz{\'e}s},
  {Sait{\={o}}}, {Takata} \& {Yamada}}{{Chamaraux}
  et~al.}{1999}]{1999MNRAS.307..236C}
{Chamaraux} P.,  {Masnou} J.-L.,  {Kaz{\'e}s} I.,  {Sait{\={o}}} M.,  {Takata}
  T.,    {Yamada} T.,  1999, \mnras, 307, 236

\bibitem[\protect\citeauthoryear{{Chiang} \& {M{\'e}nard}}{{Chiang} \&
  {M{\'e}nard}}{2019}]{chiang19}
{Chiang} Y.-K.,  {M{\'e}nard} B.,  2019, \apj, 870, 120

\bibitem[\protect\citeauthoryear{{Chilingarian}, {Melchior} \&
  {Zolotukhin}}{{Chilingarian} et~al.}{2010}]{chilingarian10}
{Chilingarian} I.~V.,  {Melchior} A.-L.,    {Zolotukhin} I.~Y.,  2010, \mnras,
  405, 1409

\bibitem[\protect\citeauthoryear{{Choloniewski} \& {Valentijn}}{{Choloniewski}
  \& {Valentijn}}{2003}]{choloniewski03}
{Choloniewski} J.,  {Valentijn} E.~A.,  2003, Acta Astron., 53, 265

\bibitem[\protect\citeauthoryear{{Collobert}, {Sarzi}, {Davies}, {Kuntschner}
  \& {Colless}}{{Collobert} et~al.}{2006}]{2006MNRAS.370.1213C}
{Collobert} M.,  {Sarzi} M.,  {Davies} R.~L.,  {Kuntschner} H.,    {Colless}
  M.,  2006, \mnras, 370, 1213

\bibitem[\protect\citeauthoryear{{Corwin} H.~G. \& {Emerson}}{{Corwin} \&
  {Emerson}}{1982}]{1982MNRAS.200..621C}
{Corwin} H.~G. J.,  {Emerson} D.,  1982, \mnras, 200, 621

\bibitem[\protect\citeauthoryear{{Cottaar}, {Covey}, {Foster}, {Meyer}, {Tan},
  {Nidever}, {Chojnowski}, {da Rio}, {Flaherty}, {Frinchaboy}, {Majewski},
  {Skrutskie}, {Wilson} \& {Zasowski}}{{Cottaar}
  et~al.}{2015}]{2015ApJ...807...27C}
{Cottaar} M.,  {Covey} K.~R.,  {Foster} J.~B.,  {Meyer} M.~R.,  {Tan} J.~C.,
  {Nidever} D.~L.,  {Chojnowski} S.~D.,  {da Rio} N.,  {Flaherty} K.~M.,
  {Frinchaboy} P.~M.,  {Majewski} S.,  {Skrutskie} M.~F.,  {Wilson} J.~C.,
  {Zasowski} G.,  2015, \apj, 807, 27

\bibitem[\protect\citeauthoryear{{Courtois} \& {Tully}}{{Courtois} \&
  {Tully}}{2015}]{2015MNRAS.447.1531C}
{Courtois} H.~M.,  {Tully} R.~B.,  2015, \mnras, 447, 1531

\bibitem[\protect\citeauthoryear{{Courtois}, {Tully}, {Fisher}, {Bonhomme},
  {Zavodny} \& {Barnes}}{{Courtois} et~al.}{2009}]{2009AJ....138.1938C}
{Courtois} H.~M.,  {Tully} R.~B.,  {Fisher} J.~R.,  {Bonhomme} N.,  {Zavodny}
  M.,    {Barnes} A.,  2009, \aj, 138, 1938

\bibitem[\protect\citeauthoryear{{Courtois}, {Tully}, {Makarov}, {Mitronova},
  {Koribalski}, {Karachentsev} \& {Fisher}}{{Courtois}
  et~al.}{2011}]{2011MNRAS.414.2005C}
{Courtois} H.~M.,  {Tully} R.~B.,  {Makarov} D.~I.,  {Mitronova} S.,
  {Koribalski} B.,  {Karachentsev} I.~D.,    {Fisher} J.~R.,  2011, \mnras,
  414, 2005

\bibitem[\protect\citeauthoryear{{Crook}, {Huchra}, {Martimbeau}, {Masters},
  {Jarrett} \& {Macri}}{{Crook} et~al.}{2007}]{2007ApJ...655..790C}
{Crook} A.~C.,  {Huchra} J.~P.,  {Martimbeau} N.,  {Masters} K.~L.,  {Jarrett}
  T.,    {Macri} L.~M.,  2007, \apj, 655, 790

\bibitem[\protect\citeauthoryear{{Crook}, {Huchra}, {Martimbeau}, {Masters},
  {Jarrett} \& {Macri}}{{Crook} et~al.}{2008}]{2008ApJ...685.1320C}
{Crook} A.~C.,  {Huchra} J.~P.,  {Martimbeau} N.,  {Masters} K.~L.,  {Jarrett}
  T.,    {Macri} L.~M.,  2008, \apj, 685, 1320

\bibitem[\protect\citeauthoryear{{Danziger} \& {Goss}}{{Danziger} \&
  {Goss}}{1983}]{1983MNRAS.202..703D}
{Danziger} I.~J.,  {Goss} W.~M.,  1983, \mnras, 202, 703

\bibitem[\protect\citeauthoryear{{Davies}, {Staveley-Smith} \&
  {Murray}}{{Davies} et~al.}{1989}]{1989MNRAS.236..171D}
{Davies} R.~D.,  {Staveley-Smith} L.,    {Murray} J.~D.,  1989, \mnras, 236,
  171

\bibitem[\protect\citeauthoryear{{Davies}, {Burstein}, {Dressler}, {Faber},
  {Lynden-Bell}, {Terlevich} \& {Wegner}}{{Davies}
  et~al.}{1987}]{1987ApJS...64..581D}
{Davies} R.~L.,  {Burstein} D.,  {Dressler} A.,  {Faber} S.~M.,  {Lynden-Bell}
  D.,  {Terlevich} R.~J.,    {Wegner} G.,  1987, \apjs, 64, 581

\bibitem[\protect\citeauthoryear{{Davoust} \& {Considere}}{{Davoust} \&
  {Considere}}{1995}]{1995AuAS..110...19D}
{Davoust} E.,  {Considere} S.,  1995, \aaps, 110, 19

\bibitem[\protect\citeauthoryear{{de Vaucouleurs}, {de Vaucouleurs}, {Corwin}
  Herold~G., {Buta}, {Paturel} \& {Fouque}}{{de Vaucouleurs}
  et~al.}{1991}]{1991RC3.9.C...0000d}
{de Vaucouleurs} G.,  {de Vaucouleurs} A.,  {Corwin} Herold~G. J.,  {Buta}
  R.~J.,  {Paturel} G.,    {Fouque} P.,  1991, {Third Reference Catalogue of
  Bright Galaxies}

\bibitem[\protect\citeauthoryear{{Demoulin}}{{Demoulin}}{1970}]{1970ApJ...160L..79D}
{Demoulin} M.-H.,  1970, \apjl, 160, L79

\bibitem[\protect\citeauthoryear{{di Nella}, {Couch}, {Parker} \&
  {Paturel}}{{di Nella} et~al.}{1997}]{1997MNRAS.287..472D}
{di Nella} H.,  {Couch} W.~J.,  {Parker} Q.~A.,    {Paturel} G.,  1997, \mnras,
  287, 472

\bibitem[\protect\citeauthoryear{{di Nella}, {Paturel}, {Walsh}, {Bottinelli},
  {Gouguenheim} \& {Theureau}}{{di Nella} et~al.}{1996}]{1996AuAS..118..311D}
{di Nella} H.,  {Paturel} G.,  {Walsh} A.~J.,  {Bottinelli} L.,  {Gouguenheim}
  L.,    {Theureau} G.,  1996, \aaps, 118, 311

\bibitem[\protect\citeauthoryear{{Djorgovski}, {Thompson}, {de Carvalho} \&
  {Mould}}{{Djorgovski} et~al.}{1990}]{1990AJ....100..599D}
{Djorgovski} S.,  {Thompson} D.~J.,  {de Carvalho} R.~R.,    {Mould} J.~R.,
  1990, \aj, 100, 599

\bibitem[\protect\citeauthoryear{{Donley}, {Koribalski}, {Staveley-Smith},
  {Kraan-Korteweg}, {Schr{\"o}der} \& {Henning}}{{Donley}
  et~al.}{2006}]{2006MNRAS.369.1741D}
{Donley} J.~L.,  {Koribalski} B.~S.,  {Staveley-Smith} L.,  {Kraan-Korteweg}
  R.~C.,  {Schr{\"o}der} A.,    {Henning} P.~A.,  2006, \mnras, 369, 1741

\bibitem[\protect\citeauthoryear{{Donley}, {Staveley-Smith}, {Kraan-Korteweg},
  {Islas-Islas}, {Schr{\"o}der}, {Henning}, {Koribalski}, {Mader} \&
  {Stewart}}{{Donley} et~al.}{2005}]{2005AJ....129..220D}
{Donley} J.~L.,  {Staveley-Smith} L.,  {Kraan-Korteweg} R.~C.,  {Islas-Islas}
  J.~M.,  {Schr{\"o}der} A.,  {Henning} P.~A.,  {Koribalski} B.,  {Mader} S.,
   {Stewart} I.,  2005, \aj, 129, 220

\bibitem[\protect\citeauthoryear{{Doyle}, {Drinkwater}, {Rohde}, {Pimbblet},
  {Read}, {Meyer}, {Zwaan}, {Ryan-Weber}, {Stevens}, {Koribalski}
  et~al.,}{{Doyle} et~al.}{2005}]{2005MNRAS.361...34D}
{Doyle} M.~T.,  {Drinkwater} M.~J.,  {Rohde} D.~J.,  {Pimbblet} K.~A.,  {Read}
  M.,  {Meyer} M.~J.,  {Zwaan} M.~A.,  {Ryan-Weber} E.,  {Stevens} J.,
  {Koribalski} B.~S.,    et~al., 2005, \mnras, 361, 34

\bibitem[\protect\citeauthoryear{{Draine} \& {Li}}{{Draine} \&
  {Li}}{2007}]{dl07}
{Draine} B.~T.,  {Li} A.,  2007, \apj, 657, 810

\bibitem[\protect\citeauthoryear{{Dressler}}{{Dressler}}{1991}]{1991ApJS...75..241D}
{Dressler} A.,  1991, \apjs, 75, 241

\bibitem[\protect\citeauthoryear{{Durret}, {Wakamatsu}, {Nagayama}, {Adami} \&
  {Biviano}}{{Durret} et~al.}{2015}]{2015AuA...583A.124D}
{Durret} F.,  {Wakamatsu} K.,  {Nagayama} T.,  {Adami} C.,    {Biviano} A.,
  2015, \aap, 583, A124

\bibitem[\protect\citeauthoryear{{Dutra}, {Ahumada}, {Clari{\'a}}, {Bica} \&
  {Barbuy}}{{Dutra} et~al.}{2003}]{dutra03}
{Dutra} C.~M.,  {Ahumada} A.~V.,  {Clari{\'a}} J.~J.,  {Bica} E.,    {Barbuy}
  B.,  2003, A\&A, 408, 287

\bibitem[\protect\citeauthoryear{{Evans}, {Kennedy}, {Dufton}, {Howarth},
  {Walborn}, {Markova}, {Clark}, {de Mink}, {de Koter} et~al.,}{{Evans}
  et~al.}{2015}]{2015AuA...574A..13E}
{Evans} C.~J.,  {Kennedy} M.~B.,  {Dufton} P.~L.,  {Howarth} I.~D.,  {Walborn}
  N.~R.,  {Markova} N.,  {Clark} J.~S.,  {de Mink} S.~E.,  {de Koter} A.,
  et~al., 2015, \aap, 574, A13

\bibitem[\protect\citeauthoryear{{Fairall}}{{Fairall}}{1981}]{1981MNRAS.196..417F}
{Fairall} A.~P.,  1981, \mnras, 196, 417

\bibitem[\protect\citeauthoryear{{Fairall}}{{Fairall}}{1983}]{1983MNRAS.203...47F}
{Fairall} A.~P.,  1983, \mnras, 203, 47

\bibitem[\protect\citeauthoryear{{Fairall}}{{Fairall}}{1988a}]{1988MNRAS.230...69F}
{Fairall} A.~P.,  1988a, \mnras, 230, 69

\bibitem[\protect\citeauthoryear{{Fairall}}{{Fairall}}{1988b}]{1988MNRAS.233..691F}
{Fairall} A.~P.,  1988b, \mnras, 233, 691

\bibitem[\protect\citeauthoryear{{Fairall}, {Vettolani} \&
  {Chincarini}}{{Fairall} et~al.}{1989}]{1989AuAS...78..269F}
{Fairall} A.~P.,  {Vettolani} G.,    {Chincarini} G.,  1989, \aaps, 78, 269

\bibitem[\protect\citeauthoryear{{Fairall} \& {Woudt}}{{Fairall} \&
  {Woudt}}{2006}]{2006MNRAS.366..267F}
{Fairall} A.~P.,  {Woudt} P.~A.,  2006, \mnras, 366, 267

\bibitem[\protect\citeauthoryear{{Fairall}, {Woudt} \&
  {Kraan-Korteweg}}{{Fairall} et~al.}{1998}]{1998AuAS..127..463F}
{Fairall} A.~P.,  {Woudt} P.~A.,    {Kraan-Korteweg} R.~C.,  1998, \aaps, 127,
  463

\bibitem[\protect\citeauthoryear{{Falco}, {Kurtz}, {Geller}, {Huchra},
  {Peters}, {Berlind}, {Mink}, {Tokarz} \& {Elwell}}{{Falco}
  et~al.}{1999}]{1999PASP..111..438F}
{Falco} E.~E.,  {Kurtz} M.~J.,  {Geller} M.~J.,  {Huchra} J.~P.,  {Peters} J.,
  {Berlind} P.,  {Mink} D.~J.,  {Tokarz} S.~P.,    {Elwell} B.,  1999, \pasp,
  111, 438

\bibitem[\protect\citeauthoryear{{F{\H{u}}r{\'e}sz}, {Hartmann}, {Megeath},
  {Szentgyorgyi} \& {Hamden}}{{F{\H{u}}r{\'e}sz}
  et~al.}{2008}]{2008ApJ...676.1109F}
{F{\H{u}}r{\'e}sz} G.,  {Hartmann} L.~W.,  {Megeath} S.~T.,  {Szentgyorgyi}
  A.~H.,    {Hamden} E.~T.,  2008, \apj, 676, 1109

\bibitem[\protect\citeauthoryear{{Fisher} \& {Tully}}{{Fisher} \&
  {Tully}}{1981}]{1981ApJS...47..139F}
{Fisher} J.~R.,  {Tully} R.~B.,  1981, \apjs, 47, 139

\bibitem[\protect\citeauthoryear{{Fisher}, {Huchra}, {Strauss}, {Davis},
  {Yahil} \& {Schlegel}}{{Fisher} et~al.}{1995}]{1995ApJS..100...69F}
{Fisher} K.~B.,  {Huchra} J.~P.,  {Strauss} M.~A.,  {Davis} M.,  {Yahil} A.,
  {Schlegel} D.,  1995, \apjs, 100, 69

\bibitem[\protect\citeauthoryear{{Fitzpatrick}}{{Fitzpatrick}}{1999}]{fitz99}
{Fitzpatrick} E.~L.,  1999, \pasp, 111, 63

\bibitem[\protect\citeauthoryear{{Foley}}{{Foley}}{2009}]{2009CBET.1855....1F}
{Foley} R.~J.,  2009, Central Bureau Electronic Telegrams, 1855, 1

\bibitem[\protect\citeauthoryear{{Foster}, {Cottaar}, {Covey}, {Arce}, {Meyer},
  {Nidever}, {Stassun}, {Tan}, {Chojnowski}, {da Rio}, {Flaherty}, {Rebull},
  {Frinchaboy}, {Majewski}, {Skrutskie}, {Wilson} \& {Zasowski}}{{Foster}
  et~al.}{2015}]{2015ApJ...799..136F}
{Foster} J.~B.,  {Cottaar} M.,  {Covey} K.~R.,  {Arce} H.~G.,  {Meyer} M.~R.,
  {Nidever} D.~L.,  {Stassun} K.~G.,  {Tan} J.~C.,  {Chojnowski} S.~D.,  {da
  Rio} N.,  {Flaherty} K.~M.,  {Rebull} L.,  {Frinchaboy} P.~M.,  {Majewski}
  S.~R.,  {Skrutskie} M.,  {Wilson} J.~C.,    {Zasowski} G.,  2015, \apj, 799,
  136

\bibitem[\protect\citeauthoryear{{Fouque}, {Durand}, {Bottinelli},
  {Gouguenheim} \& {Paturel}}{{Fouque} et~al.}{1990}]{1990AuAS...86..473F}
{Fouque} P.,  {Durand} N.,  {Bottinelli} L.,  {Gouguenheim} L.,    {Paturel}
  G.,  1990, \aaps, 86, 473

\bibitem[\protect\citeauthoryear{{Freeman}, {Karlsson}, {Lynga}, {Burrell},
  {van Woerden}, {Goss} \& {Mebold}}{{Freeman}
  et~al.}{1977}]{1977AuA....55..445F}
{Freeman} K.~C.,  {Karlsson} B.,  {Lynga} G.,  {Burrell} J.~F.,  {van Woerden}
  H.,  {Goss} W.~M.,    {Mebold} U.,  1977, \aap, 55, 445

\bibitem[\protect\citeauthoryear{{Freudling}}{{Freudling}}{1995}]{1995AuAS..112..429F}
{Freudling} W.,  1995, \aaps, 112, 429

\bibitem[\protect\citeauthoryear{{Fuerst}, {Reich}, {Kuehr} \&
  {Stickel}}{{Fuerst} et~al.}{1989}]{1989AuA...223...66F}
{Fuerst} E.,  {Reich} W.,  {Kuehr} H.,    {Stickel} M.,  1989, \aap, 223, 66

\bibitem[\protect\citeauthoryear{{Garrido}, {Marcelin}, {Amram} \&
  {Boissin}}{{Garrido} et~al.}{2003}]{2003AuA...399...51G}
{Garrido} O.,  {Marcelin} M.,  {Amram} P.,    {Boissin} O.,  2003, \aap, 399,
  51

\bibitem[\protect\citeauthoryear{{Giovanelli} \& {Haynes}}{{Giovanelli} \&
  {Haynes}}{1981}]{1981AJ.....86..340G}
{Giovanelli} R.,  {Haynes} M.~P.,  1981, \aj, 86, 340

\bibitem[\protect\citeauthoryear{{Giovanelli} \& {Haynes}}{{Giovanelli} \&
  {Haynes}}{1982}]{1982AJ.....87.1355G}
{Giovanelli} R.,  {Haynes} M.~P.,  1982, \aj, 87, 1355

\bibitem[\protect\citeauthoryear{{Goncalves}, {Martin}, {Halpern}, {Eracleous}
  \& {Pavlov}}{{Goncalves} et~al.}{2008}]{2008ATel.1623....1G}
{Goncalves} T.~S.,  {Martin} D.~C.,  {Halpern} J.~P.,  {Eracleous} M.,
  {Pavlov} G.~G.,  2008, The Astronomer's Telegram, 1623, 1

\bibitem[\protect\citeauthoryear{{Gonzalez}, {Rejkuba}, {Zoccali}, {Valenti},
  {Minniti}, {Schultheis}, {Tobar} \& {Chen}}{{Gonzalez}
  et~al.}{2012}]{gonzalez12}
{Gonzalez} O.~A.,  {Rejkuba} M.,  {Zoccali} M.,  {Valenti} E.,  {Minniti} D.,
  {Schultheis} M.,  {Tobar} R.,    {Chen} B.,  2012, \aap, 543, A13

\bibitem[\protect\citeauthoryear{{Gregory}, {Tifft}, {Moody}, {Newberry} \&
  {Hall}}{{Gregory} et~al.}{2000}]{2000AJ....119..545G}
{Gregory} S.~A.,  {Tifft} W.~G.,  {Moody} J.~W.,  {Newberry} M.~V.,    {Hall}
  S.~M.,  2000, \aj, 119, 545

\bibitem[\protect\citeauthoryear{{Harris}}{{Harris}}{1996}]{1996AJ....112.1487H}
{Harris} W.~E.,  1996, \aj, 112, 1487

\bibitem[\protect\citeauthoryear{{Hasegawa}, {Wakamatsu}, {Malkan},
  {Sekiguchi}, {Menzies}, {Parker}, {Jugaku}, {Karoji} \& {Okamura}}{{Hasegawa}
  et~al.}{2000}]{2000MNRAS.316..326H}
{Hasegawa} T.,  {Wakamatsu} K.-i.,  {Malkan} M.,  {Sekiguchi} K.,  {Menzies}
  J.~W.,  {Parker} Q.~A.,  {Jugaku} J.,  {Karoji} H.,    {Okamura} S.,  2000,
  \mnras, 316, 326

\bibitem[\protect\citeauthoryear{{Hauschildt}}{{Hauschildt}}{1987}]{1987AuA...184...43H}
{Hauschildt} M.,  1987, \aap, 184, 43

\bibitem[\protect\citeauthoryear{{Haynes}, {Giovanelli}, {Chamaraux}, {da
  Costa}, {Freudling}, {Salzer} \& {Wegner}}{{Haynes}
  et~al.}{1999}]{1999AJ....117.2039H}
{Haynes} M.~P.,  {Giovanelli} R.,  {Chamaraux} P.,  {da Costa} L.~N.,
  {Freudling} W.,  {Salzer} J.~J.,    {Wegner} G.,  1999, \aj, 117, 2039

\bibitem[\protect\citeauthoryear{{Haynes}, {Giovanelli}, {Herter}, {Vogt},
  {Freudling}, {Maia}, {Salzer} \& {Wegner}}{{Haynes}
  et~al.}{1997}]{1997AJ....113.1197H}
{Haynes} M.~P.,  {Giovanelli} R.,  {Herter} T.,  {Vogt} N.~P.,  {Freudling} W.,
   {Maia} M.~A.~G.,  {Salzer} J.~J.,    {Wegner} G.,  1997, \aj, 113, 1197

\bibitem[\protect\citeauthoryear{{Haynes}, {Giovanelli}, {Starosta} \&
  {Magri}}{{Haynes} et~al.}{1988}]{1988AJ.....95..607H}
{Haynes} M.~P.,  {Giovanelli} R.,  {Starosta} B.~M.,    {Magri} C.,  1988, \aj,
  95, 607

\bibitem[\protect\citeauthoryear{{Haynes}, {Hogg}, {Maddalena}, {Roberts} \&
  {van Zee}}{{Haynes} et~al.}{1998}]{1998AJ....115...62H}
{Haynes} M.~P.,  {Hogg} D.~E.,  {Maddalena} R.~J.,  {Roberts} M.~S.,    {van
  Zee} L.,  1998, \aj, 115, 62

\bibitem[\protect\citeauthoryear{{Henning}}{{Henning}}{1992}]{1992ApJS...78..365H}
{Henning} P.~A.,  1992, \apjs, 78, 365

\bibitem[\protect\citeauthoryear{{Henning}, {Springob}, {Minchin}, {Momjian},
  {Catinella}, {McIntyre}, {Day}, {Muller}, {Koribalski}, {Rosenberg},
  {Schneider}, {Staveley-Smith} \& {van Driel}}{{Henning}
  et~al.}{2010}]{2010AJ....139.2130H}
{Henning} P.~A.,  {Springob} C.~M.,  {Minchin} R.~F.,  {Momjian} E.,
  {Catinella} B.,  {McIntyre} T.,  {Day} F.,  {Muller} E.,  {Koribalski} B.,
  {Rosenberg} J.~L.,  {Schneider} S.,  {Staveley-Smith} L.,    {van Driel} W.,
  2010, \aj, 139, 2130

\bibitem[\protect\citeauthoryear{{Hill}, {Heasley}, {Becklin} \&
  {Wynn-Williams}}{{Hill} et~al.}{1988}]{1988AJ.....95.1031H}
{Hill} G.~J.,  {Heasley} J.~N.,  {Becklin} E.~E.,    {Wynn-Williams} C.~G.,
  1988, \aj, 95, 1031

\bibitem[\protect\citeauthoryear{{Hong}, {Staveley-Smith}, {Masters},
  {Springob}, {Macri}, {Koribalski}, {Jones}, {Jarrett} \& {Crook}}{{Hong}
  et~al.}{2013}]{2013MNRAS.432.1178H}
{Hong} T.,  {Staveley-Smith} L.,  {Masters} K.~L.,  {Springob} C.~M.,  {Macri}
  L.~M.,  {Koribalski} B.~S.,  {Jones} D.~H.,  {Jarrett} T.~H.,    {Crook}
  A.~C.,  2013, \mnras, 432, 1178

\bibitem[\protect\citeauthoryear{{Huchra}, {Davis}, {Latham} \&
  {Tonry}}{{Huchra} et~al.}{1983}]{1983ApJS...52...89H}
{Huchra} J.,  {Davis} M.,  {Latham} D.,    {Tonry} J.,  1983, \apjs, 52, 89

\bibitem[\protect\citeauthoryear{{Huchra}, {Geller}, {Clemens}, {Tokarz} \&
  {Michel}}{{Huchra} et~al.}{1995}]{1995--CfAzcat--H}
{Huchra} J.,  {Geller} M.,  {Clemens} C.,  {Tokarz} S.,    {Michel} A.,  1995

\bibitem[\protect\citeauthoryear{{Huchra}, {Macri}, {Masters}, {Jarrett},
  {Berlind} \& {Calkins}}{{Huchra} et~al.}{2012}]{huchra12}
{Huchra} J.~P.,  {Macri} L.~M.,  {Masters} K.~L.,  {Jarrett} T.~H.,  {Berlind}
  P.,    {Calkins} M.,  2012, ApJS, 199, 26

\bibitem[\protect\citeauthoryear{{Huchra}, {Macri}, {Masters}, {Jarrett},
  {Berlind}, {Calkins}, {Crook}, {Cutri}, {Erdo{\v{g}}du} et~al.,}{{Huchra}
  et~al.}{2012}]{2012ApJS..199...26H}
{Huchra} J.~P.,  {Macri} L.~M.,  {Masters} K.~L.,  {Jarrett} T.~H.,  {Berlind}
  P.,  {Calkins} M.,  {Crook} A.~C.,  {Cutri} R.,  {Erdo{\v{g}}du} P.,
  et~al., 2012, \apjs, 199, 26

\bibitem[\protect\citeauthoryear{{Huchra}, {Vogeley} \& {Geller}}{{Huchra}
  et~al.}{1999}]{1999ApJS..121..287H}
{Huchra} J.~P.,  {Vogeley} M.~S.,    {Geller} M.~J.,  1999, \apjs, 121, 287

\bibitem[\protect\citeauthoryear{{Huchtmeier}, {Karachentsev} \&
  {Karachentseva}}{{Huchtmeier} et~al.}{2001}]{2001AuA...377..801H}
{Huchtmeier} W.~K.,  {Karachentsev} I.~D.,    {Karachentseva} V.~E.,  2001,
  \aap, 377, 801

\bibitem[\protect\citeauthoryear{{Huchtmeier}, {Karachentsev} \&
  {Karachentseva}}{{Huchtmeier} et~al.}{2003}]{2003AuA...401..483H}
{Huchtmeier} W.~K.,  {Karachentsev} I.~D.,    {Karachentseva} V.~E.,  2003,
  \aap, 401, 483

\bibitem[\protect\citeauthoryear{{Huchtmeier}, {Karachentsev}, {Karachentseva}
  \& {Ehle}}{{Huchtmeier} et~al.}{2000}]{2000AuAS..141..469H}
{Huchtmeier} W.~K.,  {Karachentsev} I.~D.,  {Karachentseva} V.~E.,    {Ehle}
  M.,  2000, \aaps, 141, 469

\bibitem[\protect\citeauthoryear{{Huchtmeier}, {Karachentsev}, {Karachentseva},
  {Kudrya} \& {Mitronova}}{{Huchtmeier} et~al.}{2005}]{2005AuA...435..459H}
{Huchtmeier} W.~K.,  {Karachentsev} I.~D.,  {Karachentseva} V.~E.,  {Kudrya}
  Y.~N.,    {Mitronova} S.~N.,  2005, \aap, 435, 459

\bibitem[\protect\citeauthoryear{{Huchtmeier}, {Lercher}, {Seeberger}, {Saurer}
  \& {Weinberger}}{{Huchtmeier} et~al.}{1995}]{1995AuA...293L..33H}
{Huchtmeier} W.~K.,  {Lercher} G.,  {Seeberger} R.,  {Saurer} W.,
  {Weinberger} R.,  1995, \aap, 293, L33

\bibitem[\protect\citeauthoryear{{Huchtmeier} \& {Richter}}{{Huchtmeier} \&
  {Richter}}{1989}]{1989gcho.book.....H}
{Huchtmeier} W.~K.,  {Richter} O.~G.,  1989, {A General Catalog of HI
  Observations of Galaxies. The Reference Catalog.}

\bibitem[\protect\citeauthoryear{{Humason}, {Mayall} \& {Sandage}}{{Humason}
  et~al.}{1956}]{1956AJ.....61...97H}
{Humason} M.~L.,  {Mayall} N.~U.,    {Sandage} A.~R.,  1956, \aj, 61, 97

\bibitem[\protect\citeauthoryear{{Humason} \& {Wahlquist}}{{Humason} \&
  {Wahlquist}}{1955}]{1955AJ.....60..254H}
{Humason} M.~L.,  {Wahlquist} H.~D.,  1955, \aj, 60, 254

\bibitem[\protect\citeauthoryear{{Im}, {Lee}, {Cho}, {Choi}, {Ko} \&
  {Song}}{{Im} et~al.}{2007}]{2007ApJ...664...64I}
{Im} M.,  {Lee} I.,  {Cho} Y.,  {Choi} C.,  {Ko} J.,    {Song} M.,  2007, \apj,
  664, 64

\bibitem[\protect\citeauthoryear{{Jarrett}}{{Jarrett}}{2004}]{jarrett04}
{Jarrett} T.~H.,  2004, Publications of the Astronomical Society of Australia,
  21, 396

\bibitem[\protect\citeauthoryear{{Jarrett}, {Chester}, {Cutri}, {Schneider},
  {Rosenberg}, {Huchra} \& {Mader}}{{Jarrett}
  et~al.}{2000}]{2000AJ....120..298J}
{Jarrett} T.~H.,  {Chester} T.,  {Cutri} R.,  {Schneider} S.,  {Rosenberg} J.,
  {Huchra} J.~P.,    {Mader} J.,  2000, \aj, 120, 298

\bibitem[\protect\citeauthoryear{{Jones}, {Read}, {Saunders}, {Colless},
  {Jarrett}, {Parker}, {Fairall}, {Mauch}, {Sadler} et~al.,}{{Jones}
  et~al.}{2009}]{2009MNRAS.399..683J}
{Jones} D.~H.,  {Read} M.~A.,  {Saunders} W.,  {Colless} M.,  {Jarrett} T.,
  {Parker} Q.~A.,  {Fairall} A.~P.,  {Mauch} T.,  {Sadler} E.~M.,    et~al.,
  2009, \mnras, 399, 683

\bibitem[\protect\citeauthoryear{{Jones}, {Saunders}, {Read} \&
  {Colless}}{{Jones} et~al.}{2005}]{2005PASA...22..277J}
{Jones} D.~H.,  {Saunders} W.,  {Read} M.,    {Colless} M.,  2005, \pasa, 22,
  277

\bibitem[\protect\citeauthoryear{{Jones} \& {McAdam}}{{Jones} \&
  {McAdam}}{1992}]{1992ApJS...80..137J}
{Jones} P.~A.,  {McAdam} W.~B.,  1992, \apjs, 80, 137

\bibitem[\protect\citeauthoryear{{Juraszek}, {Staveley-Smith},
  {Kraan-Korteweg}, {Green}, {Ekers}, {Haynes}, {Henning}, {Kesteven},
  {Koribalski}, {Price}, {Sadler} \& {Schr{\"o}der}}{{Juraszek}
  et~al.}{2000}]{2000AJ....119.1627J}
{Juraszek} S.~J.,  {Staveley-Smith} L.,  {Kraan-Korteweg} R.~C.,  {Green}
  A.~J.,  {Ekers} R.~D.,  {Haynes} R.~F.,  {Henning} P.~A.,  {Kesteven} M.~J.,
  {Koribalski} B.,  {Price} R.~M.,  {Sadler} E.~M.,    {Schr{\"o}der} A.,
  2000, \aj, 119, 1627

\bibitem[\protect\citeauthoryear{{Karachentsev}}{{Karachentsev}}{1980}]{1980ApJS...44..137K}
{Karachentsev} I.~D.,  1980, \apjs, 44, 137

\bibitem[\protect\citeauthoryear{{Karachentseva}, {Mitronova}, {Melnyk} \&
  {Karachentsev}}{{Karachentseva} et~al.}{2010}]{2010AstBu..65....1K}
{Karachentseva} V.~E.,  {Mitronova} S.~N.,  {Melnyk} O.~V.,    {Karachentsev}
  I.~D.,  2010, Astrophysical Bulletin, 65, 1

\bibitem[\protect\citeauthoryear{{Kerr} \& {Henning}}{{Kerr} \&
  {Henning}}{1987}]{1987ApJ...320L..99K}
{Kerr} F.~J.,  {Henning} P.~A.,  1987, \apjl, 320, L99

\bibitem[\protect\citeauthoryear{{Kharchenko}, {Piskunov}, {Schilbach},
  {R{\"o}ser} \& {Scholz}}{{Kharchenko} et~al.}{2013}]{2013AuA...558A..53K}
{Kharchenko} N.~V.,  {Piskunov} A.~E.,  {Schilbach} E.,  {R{\"o}ser} S.,
  {Scholz} R.~D.,  2013, \aap, 558, A53

\bibitem[\protect\citeauthoryear{{Kilborn}, {Webster}, {Staveley-Smith},
  {Marquarding}, {Banks}, {Barnes}, {Bhathal}, {de Blok}, {Boyce}
  et~al.,}{{Kilborn} et~al.}{2002}]{2002AJ....124..690K}
{Kilborn} V.~A.,  {Webster} R.~L.,  {Staveley-Smith} L.,  {Marquarding} M.,
  {Banks} G.~D.,  {Barnes} D.~G.,  {Bhathal} R.,  {de Blok} W.~J.~G.,  {Boyce}
  P.~J.,    et~al., 2002, \aj, 124, 690

\bibitem[\protect\citeauthoryear{{Kirhakos} \& {Steiner}}{{Kirhakos} \&
  {Steiner}}{1990}]{1990AJ.....99.1722K}
{Kirhakos} S.~D.,  {Steiner} J.~E.,  1990, \aj, 99, 1722

\bibitem[\protect\citeauthoryear{{Kobulnicky}, {Dickey}, {Sargent}, {Hogg} \&
  {Conti}}{{Kobulnicky} et~al.}{1995}]{1995AJ....110..116K}
{Kobulnicky} H.~A.,  {Dickey} J.~M.,  {Sargent} A.~I.,  {Hogg} D.~E.,
  {Conti} P.~S.,  1995, \aj, 110, 116

\bibitem[\protect\citeauthoryear{{Koribalski}, {Staveley-Smith}, {Kilborn},
  {Ryder}, {Kraan-Korteweg}, {Ryan-Weber}, {Ekers}, {Jerjen}, {Henning}
  et~al.,}{{Koribalski} et~al.}{2004}]{2004AJ....128...16K}
{Koribalski} B.~S.,  {Staveley-Smith} L.,  {Kilborn} V.~A.,  {Ryder} S.~D.,
  {Kraan-Korteweg} R.~C.,  {Ryan-Weber} E.~V.,  {Ekers} R.~D.,  {Jerjen} H.,
  {Henning} P.~A.,    et~al., 2004, \aj, 128, 16

\bibitem[\protect\citeauthoryear{{Kraan-Korteweg}, {Fairall} \&
  {Balkowski}}{{Kraan-Korteweg} et~al.}{1995}]{1995AuA...297..617K}
{Kraan-Korteweg} R.~C.,  {Fairall} A.~P.,    {Balkowski} C.,  1995, \aap, 297,
  617

\bibitem[\protect\citeauthoryear{{Kraan-Korteweg}, {Henning} \&
  {Schr{\"o}der}}{{Kraan-Korteweg} et~al.}{2002}]{2002AuA...391..887K}
{Kraan-Korteweg} R.~C.,  {Henning} P.~A.,    {Schr{\"o}der} A.~C.,  2002, \aap,
  391, 887

\bibitem[\protect\citeauthoryear{{Kraan-Korteweg} \&
  {Huchtmeier}}{{Kraan-Korteweg} \& {Huchtmeier}}{1992}]{1992AuA...266..150K}
{Kraan-Korteweg} R.~C.,  {Huchtmeier} W.~K.,  1992, \aap, 266, 150

\bibitem[\protect\citeauthoryear{{Kraan-Korteweg}, {van Driel}, {Schr{\"o}der},
  {Ramatsoku} \& {Henning}}{{Kraan-Korteweg}
  et~al.}{2018}]{2018MNRAS.481.1262K}
{Kraan-Korteweg} R.~C.,  {van Driel} W.,  {Schr{\"o}der} A.~C.,  {Ramatsoku}
  M.,    {Henning} P.~A.,  2018, \mnras, 481, 1262

\bibitem[\protect\citeauthoryear{{Kregel}, {van der Kruit} \& {de
  Blok}}{{Kregel} et~al.}{2004}]{2004MNRAS.352..768K}
{Kregel} M.,  {van der Kruit} P.~C.,    {de Blok} W.~J.~G.,  2004, \mnras, 352,
  768

\bibitem[\protect\citeauthoryear{{Lahav}, {Brosch}, {Goldberg}, {Hau},
  {Kraan-Korteweg} \& {Loan}}{{Lahav} et~al.}{1998}]{1998MNRAS.299...24L}
{Lahav} O.,  {Brosch} N.,  {Goldberg} E.,  {Hau} G. K.~T.,  {Kraan-Korteweg}
  R.~C.,    {Loan} A.~J.,  1998, \mnras, 299, 24

\bibitem[\protect\citeauthoryear{{Lang}, {Boyce}, {Kilborn}, {Minchin},
  {Disney}, {Jordan}, {Grossi}, {Garcia}, {Freeman}, {Phillipps} \&
  {Wright}}{{Lang} et~al.}{2003}]{2003MNRAS.342..738L}
{Lang} R.~H.,  {Boyce} P.~J.,  {Kilborn} V.~A.,  {Minchin} R.~F.,  {Disney}
  M.~J.,  {Jordan} C.~A.,  {Grossi} M.,  {Garcia} D.~A.,  {Freeman} K.~C.,
  {Phillipps} S.,    {Wright} A.~E.,  2003, \mnras, 342, 738

\bibitem[\protect\citeauthoryear{{Lara}, {M{\'a}rquez}, {Cotton}, {Feretti},
  {Giovannini}, {Marcaide} \& {Venturi}}{{Lara}
  et~al.}{2001}]{2001AuA...378..826L}
{Lara} L.,  {M{\'a}rquez} I.,  {Cotton} W.~D.,  {Feretti} L.,  {Giovannini} G.,
   {Marcaide} J.~M.,    {Venturi} T.,  2001, \aap, 378, 826

\bibitem[\protect\citeauthoryear{{Lavaux} \& {Hudson}}{{Lavaux} \&
  {Hudson}}{2011}]{2011MNRAS.416.2840L}
{Lavaux} G.,  {Hudson} M.~J.,  2011, \mnras, 416, 2840

\bibitem[\protect\citeauthoryear{{Lawrence}, {Rowan-Robinson}, {Ellis},
  {Frenk}, {Efstathiou}, {Kaiser}, {Saunders}, {Parry}, {Xiaoyang} \&
  {Crawford}}{{Lawrence} et~al.}{1999}]{1999MNRAS.308..897L}
{Lawrence} A.,  {Rowan-Robinson} M.,  {Ellis} R.~S.,  {Frenk} C.~S.,
  {Efstathiou} G.,  {Kaiser} N.,  {Saunders} W.,  {Parry} I.~R.,  {Xiaoyang}
  X.,    {Crawford} J.,  1999, \mnras, 308, 897

\bibitem[\protect\citeauthoryear{{Lenz}, {Hensley} \& {Dor{\'e}}}{{Lenz}
  et~al.}{2017}]{lenz17}
{Lenz} D.,  {Hensley} B.~S.,    {Dor{\'e}} O.,  2017, \apj, 846, 38

\bibitem[\protect\citeauthoryear{{Lewis}}{{Lewis}}{1983}]{1983AJ.....88..962L}
{Lewis} B.~M.,  1983, \aj, 88, 962

\bibitem[\protect\citeauthoryear{{Lewis}}{{Lewis}}{1987}]{1987ApJS...63..515L}
{Lewis} B.~M.,  1987, \apjs, 63, 515

\bibitem[\protect\citeauthoryear{{Lewis} \& {Davies}}{{Lewis} \&
  {Davies}}{1973}]{1973MNRAS.165..213L}
{Lewis} B.~M.,  {Davies} R.~D.,  1973, \mnras, 165, 213

\bibitem[\protect\citeauthoryear{{Lewis}, {Helou} \& {Salpeter}}{{Lewis}
  et~al.}{1985}]{1985ApJS...59..161L}
{Lewis} B.~M.,  {Helou} G.,    {Salpeter} E.~E.,  1985, \apjs, 59, 161

\bibitem[\protect\citeauthoryear{{Liu}, {Zhao} \& {Hou}}{{Liu}
  et~al.}{2015}]{2015RAA....15.1089L}
{Liu} X.-W.,  {Zhao} G.,    {Hou} J.-L.,  2015, Research in Astronomy and
  Astrophysics, 15, 1089

\bibitem[\protect\citeauthoryear{{Longmore}, {Hawarden}, {Goss}, {Mebold} \&
  {Webster}}{{Longmore} et~al.}{1982}]{1982MNRAS.200..325L}
{Longmore} A.~J.,  {Hawarden} T.~G.,  {Goss} W.~M.,  {Mebold} U.,    {Webster}
  B.~L.,  1982, \mnras, 200, 325

\bibitem[\protect\citeauthoryear{{Lu}, {Dow}, {Houck}, {Salpeter} \&
  {Lewis}}{{Lu} et~al.}{1990}]{1990ApJ...357..388L}
{Lu} N.~Y.,  {Dow} M.~W.,  {Houck} J.~R.,  {Salpeter} E.~E.,    {Lewis} B.~M.,
  1990, \apj, 357, 388

\bibitem[\protect\citeauthoryear{{Lucas}, {Hoare}, {Longmore}, {Schr{\"o}der},
  {Davis}, {Adamson}, {Bandyopadhyay}, {de Grijs}, {Smith}, {Gosling},
  {Mitchison}, {G{\'a}sp{\'a}r}, {Coe}, {Tamura}, {Parker} et~al.,}{{Lucas}
  et~al.}{2008}]{lucas08}
{Lucas} P.~W.,  {Hoare} M.~G.,  {Longmore} A.,  {Schr{\"o}der} A.~C.,  {Davis}
  C.~J.,  {Adamson} A.,  {Bandyopadhyay} R.~M.,  {de Grijs} R.,  {Smith} M.,
  {Gosling} A.,  {Mitchison} S.,  {G{\'a}sp{\'a}r} A.,  {Coe} M.,  {Tamura} M.,
   {Parker} Q.,    et~al., 2008, \mnras, 391, 136

\bibitem[\protect\citeauthoryear{{Macri}, {Kraan-Korteweg}, {Lambert},
  {Alonso}, {Berlind}, {Calkins}, {Erdo{\u{g}}du}, {Falco}, {Jarrett} \&
  {Mink}}{{Macri} et~al.}{2019}]{macri19}
{Macri} L.~M.,  {Kraan-Korteweg} R.~C.,  {Lambert} T.,  {Alonso} M.~V.,
  {Berlind} P.,  {Calkins} M.,  {Erdo{\u{g}}du} P.,  {Falco} E.~E.,  {Jarrett}
  T.~H.,    {Mink} J.~D.,  2019, \apjs, 245, 6

\bibitem[\protect\citeauthoryear{{Makarov}, {Karachentsev} \&
  {Burenkov}}{{Makarov} et~al.}{2003}]{2003AuA...405..951M}
{Makarov} D.~I.,  {Karachentsev} I.~D.,    {Burenkov} A.~N.,  2003, \aap, 405,
  951

\bibitem[\protect\citeauthoryear{{Makarov}, {Karachentsev}, {Burennkov},
  {Tyurina} \& {Korotkova}}{{Makarov} et~al.}{1997}]{1997AstL...23..638M}
{Makarov} D.~I.,  {Karachentsev} I.~D.,  {Burennkov} A.~N.,  {Tyurina} N.~V.,
   {Korotkova} G.~G.,  1997, Astronomy Letters, 23, 638

\bibitem[\protect\citeauthoryear{{Martin}, {Bottinelli}, {Dennefeld}, {Fouque},
  {Gouguenheim} \& {Paturel}}{{Martin} et~al.}{1990}]{1990AuA...235...41M}
{Martin} J.~M.,  {Bottinelli} L.,  {Dennefeld} M.,  {Fouque} P.,  {Gouguenheim}
  L.,    {Paturel} G.,  1990, \aap, 235, 41

\bibitem[\protect\citeauthoryear{{Martin}, {Bottinelli}, {Dennefeld} \&
  {Gouguenheim}}{{Martin} et~al.}{1991}]{1991AuA...245..393M}
{Martin} J.~M.,  {Bottinelli} L.,  {Dennefeld} M.,    {Gouguenheim} L.,  1991,
  \aap, 245, 393

\bibitem[\protect\citeauthoryear{{Marzke}, {Huchra} \& {Geller}}{{Marzke}
  et~al.}{1996}]{1996AJ....112.1803M}
{Marzke} R.~O.,  {Huchra} J.~P.,    {Geller} M.~J.,  1996, \aj, 112, 1803

\bibitem[\protect\citeauthoryear{{Masetti}, {Mason}, {Morelli}, {Cellone},
  {McBride}, {Palazzi}, {Bassani}, {Bazzano}, {Bird}, {Charles}, {Dean},
  {Galaz} et~al.,}{{Masetti} et~al.}{2008}]{2008AuA...482..113M}
{Masetti} N.,  {Mason} E.,  {Morelli} L.,  {Cellone} S.~A.,  {McBride} V.~A.,
  {Palazzi} E.,  {Bassani} L.,  {Bazzano} A.,  {Bird} A.~J.,  {Charles} P.~A.,
  {Dean} A.~J.,  {Galaz} G.,    et~al., 2008, \aap, 482, 113

\bibitem[\protect\citeauthoryear{{Masetti}, {Morelli}, {Palazzi}, {Galaz},
  {Bassani}, {Bazzano}, {Bird}, {Dean}, {Israel}, {Landi}, {Malizia}, {Minniti}
  et~al.,}{{Masetti} et~al.}{2006}]{2006AuA...459...21M}
{Masetti} N.,  {Morelli} L.,  {Palazzi} E.,  {Galaz} G.,  {Bassani} L.,
  {Bazzano} A.,  {Bird} A.~J.,  {Dean} A.~J.,  {Israel} G.~L.,  {Landi} R.,
  {Malizia} A.,  {Minniti} D.,    et~al., 2006, \aap, 459, 21

\bibitem[\protect\citeauthoryear{{Masetti}, {Parisi}, {Palazzi},
  {Jim{\'e}nez-Bail{\'o}n}, {Chavushyan}, {McBride}, {Rojas}, {Steward},
  {Bassani}, {Bazzano} et~al.,}{{Masetti} et~al.}{2013}]{2013AuA...556A.120M}
{Masetti} N.,  {Parisi} P.,  {Palazzi} E.,  {Jim{\'e}nez-Bail{\'o}n} E.,
  {Chavushyan} V.,  {McBride} V.,  {Rojas} A.~F.,  {Steward} L.,  {Bassani} L.,
   {Bazzano} A.,    et~al., 2013, \aap, 556, A120

\bibitem[\protect\citeauthoryear{{Masters}, {Crook}, {Hong}, {Jarrett},
  {Koribalski}, {Macri}, {Springob} \& {Staveley-Smith}}{{Masters}
  et~al.}{2014}]{2014MNRAS.443.1044M}
{Masters} K.~L.,  {Crook} A.,  {Hong} T.,  {Jarrett} T.~H.,  {Koribalski}
  B.~S.,  {Macri} L.,  {Springob} C.~M.,    {Staveley-Smith} L.,  2014, \mnras,
  443, 1044

\bibitem[\protect\citeauthoryear{{Mathewson} \& {Ford}}{{Mathewson} \&
  {Ford}}{1996}]{1996ApJS..107...97M}
{Mathewson} D.~S.,  {Ford} V.~L.,  1996, \apjs, 107, 97

\bibitem[\protect\citeauthoryear{{Mauch} \& {Sadler}}{{Mauch} \&
  {Sadler}}{2007}]{2007MNRAS.375..931M}
{Mauch} T.,  {Sadler} E.~M.,  2007, \mnras, 375, 931

\bibitem[\protect\citeauthoryear{{McGaugh}, {Rubin} \& {de Blok}}{{McGaugh}
  et~al.}{2001}]{2001AJ....122.2381M}
{McGaugh} S.~S.,  {Rubin} V.~C.,    {de Blok} W.~J.~G.,  2001, \aj, 122, 2381

\bibitem[\protect\citeauthoryear{{McIntyre}, {Henning}, {Minchin}, {Momjian} \&
  {Butcher}}{{McIntyre} et~al.}{2015}]{2015AJ....150...28M}
{McIntyre} T.~P.,  {Henning} P.~A.,  {Minchin} R.~F.,  {Momjian} E.,
  {Butcher} Z.,  2015, \aj, 150, 28

\bibitem[\protect\citeauthoryear{{Meisner} \& {Finkbeiner}}{{Meisner} \&
  {Finkbeiner}}{2015}]{meisner15}
{Meisner} A.~M.,  {Finkbeiner} D.~P.,  2015, \apj, 798, 88

\bibitem[\protect\citeauthoryear{{Meyer}, {Zwaan}, {Webster}, {Staveley-Smith},
  {Ryan-Weber}, {Drinkwater}, {Barnes}, {Howlett}, {Kilborn} et~al.,}{{Meyer}
  et~al.}{2004}]{2004MNRAS.350.1195M}
{Meyer} M.~J.,  {Zwaan} M.~A.,  {Webster} R.~L.,  {Staveley-Smith} L.,
  {Ryan-Weber} E.,  {Drinkwater} M.~J.,  {Barnes} D.~G.,  {Howlett} M.,
  {Kilborn} V.~A.,    et~al., 2004, \mnras, 350, 1195

\bibitem[\protect\citeauthoryear{{Michel} \& {Huchra}}{{Michel} \&
  {Huchra}}{1988}]{1988PASP..100.1423M}
{Michel} A.,  {Huchra} J.,  1988, \pasp, 100, 1423

\bibitem[\protect\citeauthoryear{{Minniti}, {Lucas}, {Emerson}, {Saito},
  {Hempel}, {Pietrukowicz}, {Ahumada}, {Alonso}, {Alonso-Garcia}, {Arias},
  {Bandyopadhyay}, {Barb{\'a}}, {Barbuy}, {Bedin}, {Bica}, {Borissova},
  {Bronfman} et~al.,}{{Minniti} et~al.}{2010}]{minniti10}
{Minniti} D.,  {Lucas} P.~W.,  {Emerson} J.~P.,  {Saito} R.~K.,  {Hempel} M.,
  {Pietrukowicz} P.,  {Ahumada} A.~V.,  {Alonso} M.~V.,  {Alonso-Garcia} J.,
  {Arias} J.~I.,  {Bandyopadhyay} R.~M.,  {Barb{\'a}} R.~H.,  {Barbuy} B.,
  {Bedin} L.~R.,  {Bica} E.,  {Borissova} J.,  {Bronfman} L.,    et~al., 2010,
  New Astronomy, 15, 433

\bibitem[\protect\citeauthoryear{{Mitronova}, {Huchtmeier}, {Karachentsev},
  {Karachentseva} \& {Kudrya}}{{Mitronova} et~al.}{2005}]{2005AstL...31..501M}
{Mitronova} S.~N.,  {Huchtmeier} W.~K.,  {Karachentsev} I.~D.,  {Karachentseva}
  V.~E.,    {Kudrya} Y.~N.,  2005, Astronomy Letters, 31, 501

\bibitem[\protect\citeauthoryear{{Morganti}, {Emonts} \&
  {Oosterloo}}{{Morganti} et~al.}{2009}]{2009AuA...496L...9M}
{Morganti} R.,  {Emonts} B.,    {Oosterloo} T.,  2009, \aap, 496, L9

\bibitem[\protect\citeauthoryear{{Mould}, {Staveley-Smith}, {Schommer},
  {Bothun}, {Hall}, {Han}, {Huchra}, {Roth}, {Walsh} \& {Wright}}{{Mould}
  et~al.}{1991}]{1991ApJ...383..467M}
{Mould} J.~R.,  {Staveley-Smith} L.,  {Schommer} R.~A.,  {Bothun} G.~D.,
  {Hall} P.~J.,  {Han} M.~S.,  {Huchra} J.~P.,  {Roth} J.,  {Walsh} W.,
  {Wright} A.~E.,  1991, \apj, 383, 467

\bibitem[\protect\citeauthoryear{{Nagayama}, {Woudt}, {Nagashima}, {Nakajima},
  {Kato}, {Kurita}, {Nagata}, {Nakaya}, {Tamura}, {Sugitani}, {Wakamatsu} \&
  {Sato}}{{Nagayama} et~al.}{2004}]{nagayama04}
{Nagayama} T.,  {Woudt} P.~A.,  {Nagashima} C.,  {Nakajima} Y.,  {Kato} D.,
  {Kurita} M.,  {Nagata} T.,  {Nakaya} H.,  {Tamura} M.,  {Sugitani} K.,
  {Wakamatsu} K.,    {Sato} S.,  2004, MNRAS, 354, 980

\bibitem[\protect\citeauthoryear{{Nakanishi}, {Takata}, {Yamada}, {Takeuchi},
  {Shiroya}, {Miyazawa}, {Watanabe} \& {Sait{\={o}}}}{{Nakanishi}
  et~al.}{1997}]{1997ApJS..112..245N}
{Nakanishi} K.,  {Takata} T.,  {Yamada} T.,  {Takeuchi} T.~T.,  {Shiroya} R.,
  {Miyazawa} M.,  {Watanabe} S.,    {Sait{\={o}}} M.,  1997, \apjs, 112, 245

\bibitem[\protect\citeauthoryear{{Nataf}, {Gould}, {Fouqu{\'e}}, {Gonzalez},
  {Johnson}, {Skowron}, {Udalski}, {Szyma{\'n}ski}, {Kubiak},
  {Pietrzy{\'n}ski}, {Soszy{\'n}ski}, {Ulaczyk}, {Wyrzykowski} \&
  {Poleski}}{{Nataf} et~al.}{2013}]{nataf13}
{Nataf} D.~M.,  {Gould} A.,  {Fouqu{\'e}} P.,  {Gonzalez} O.~A.,  {Johnson}
  J.~A.,  {Skowron} J.,  {Udalski} A.,  {Szyma{\'n}ski} M.~K.,  {Kubiak} M.,
  {Pietrzy{\'n}ski} G.,  {Soszy{\'n}ski} I.,  {Ulaczyk} K.,  {Wyrzykowski}
  {\L}.,    {Poleski} R.,  2013, \apj, 769, 88

\bibitem[\protect\citeauthoryear{{Nidever}, {Zasowski} \& {Majewski}}{{Nidever}
  et~al.}{2012}]{nidever12}
{Nidever} D.~L.,  {Zasowski} G.,    {Majewski} S.~R.,  2012, \apjs, 201, 35

\bibitem[\protect\citeauthoryear{{Ogando}, {Maia}, {Pellegrini} \& {da
  Costa}}{{Ogando} et~al.}{2008}]{2008AJ....135.2424O}
{Ogando} R. L.~C.,  {Maia} M. A.~G.,  {Pellegrini} P.~S.,    {da Costa} L.~N.,
  2008, \aj, 135, 2424

\bibitem[\protect\citeauthoryear{{Oh}, {Hunter}, {Brinks}, {Elmegreen},
  {Schruba}, {Walter}, {Rupen}, {Young}, {Simpson}, {Johnson}, {Herrmann},
  {Ficut-Vicas}, {Cigan}, {Heesen}, {Ashley} \& {Zhang}}{{Oh}
  et~al.}{2015}]{2015AJ....149..180O}
{Oh} S.-H.,  {Hunter} D.~A.,  {Brinks} E.,  {Elmegreen} B.~G.,  {Schruba} A.,
  {Walter} F.,  {Rupen} M.~P.,  {Young} L.~M.,  {Simpson} C.~E.,  {Johnson}
  M.~C.,  {Herrmann} K.~A.,  {Ficut-Vicas} D.,  {Cigan} P.,  {Heesen} V.,
  {Ashley} T.,    {Zhang} H.-X.,  2015, \aj, 149, 180

\bibitem[\protect\citeauthoryear{{Owen}, {Ledlow}, {Morrison} \& {Hill}}{{Owen}
  et~al.}{1997}]{1997ApJ...488L..15O}
{Owen} F.~N.,  {Ledlow} M.~J.,  {Morrison} G.~E.,    {Hill} J.~M.,  1997,
  \apjl, 488, L15

\bibitem[\protect\citeauthoryear{{Pantoja}, {Altschuler}, {Giovanardi} \&
  {Giovanelli}}{{Pantoja} et~al.}{1997}]{1997AJ....113..905P}
{Pantoja} C.~A.,  {Altschuler} D.~R.,  {Giovanardi} C.,    {Giovanelli} R.,
  1997, \aj, 113, 905

\bibitem[\protect\citeauthoryear{{Pantoja}, {Giovanardi}, {Altschuler} \&
  {Giovanelli}}{{Pantoja} et~al.}{1994}]{1994AJ....108..921P}
{Pantoja} C.~A.,  {Giovanardi} C.,  {Altschuler} D.~R.,    {Giovanelli} R.,
  1994, \aj, 108, 921

\bibitem[\protect\citeauthoryear{{Parisi}, {Masetti}, {Jim{\'e}nez-Bail{\'o}n},
  {Chavushyan}, {Malizia}, {Landi}, {Molina}, {Fiocchi}, {Palazzi}, {Bassani},
  {Bazzano} et~al.,}{{Parisi} et~al.}{2009}]{2009AuA...507.1345P}
{Parisi} P.,  {Masetti} N.,  {Jim{\'e}nez-Bail{\'o}n} E.,  {Chavushyan} V.,
  {Malizia} A.,  {Landi} R.,  {Molina} M.,  {Fiocchi} M.,  {Palazzi} E.,
  {Bassani} L.,  {Bazzano} A.,    et~al., 2009, \aap, 507, 1345

\bibitem[\protect\citeauthoryear{{Paturel}, {Theureau}, {Bottinelli},
  {Gouguenheim}, {Coudreau-Durand}, {Hallet} \& {Petit}}{{Paturel}
  et~al.}{2003}]{2003AuA...412...57P}
{Paturel} G.,  {Theureau} G.,  {Bottinelli} L.,  {Gouguenheim} L.,
  {Coudreau-Durand} N.,  {Hallet} N.,    {Petit} C.,  2003, \aap, 412, 57

\bibitem[\protect\citeauthoryear{{Perlman}, {Stocke}, {Schachter}, {Elvis},
  {Ellingson}, {Urry}, {Potter}, {Impey} \& {Kolchinsky}}{{Perlman}
  et~al.}{1996}]{1996ApJS..104..251P}
{Perlman} E.~S.,  {Stocke} J.~T.,  {Schachter} J.~F.,  {Elvis} M.,  {Ellingson}
  E.,  {Urry} C.~M.,  {Potter} M.,  {Impey} C.~D.,    {Kolchinsky} P.,  1996,
  \apjs, 104, 251

\bibitem[\protect\citeauthoryear{{Pfleiderer}, {Gruber}, {Gruber} \&
  {Velden}}{{Pfleiderer} et~al.}{1981}]{1981AuA...102L..21P}
{Pfleiderer} J.,  {Gruber} M.~D.,  {Gruber} G.~M.,    {Velden} L.,  1981, \aap,
  102, L21

\bibitem[\protect\citeauthoryear{{Planck Collaboration}, {Abergel}, {Ade},
  {Aghanim}, {Alves}, {Aniano}, {Armitage-Caplan}, {Arnaud}, {Ashdown},
  {Atrio-Barand ela}, {Aumont}, {Baccigalupi} et~al.,}{{Planck Collaboration}
  et~al.}{2014}]{planck14}
{Planck Collaboration} {Abergel} A.,  {Ade} P.~A.~R.,  {Aghanim} N.,  {Alves}
  M.~I.~R.,  {Aniano} G.,  {Armitage-Caplan} C.,  {Arnaud} M.,  {Ashdown} M.,
  {Atrio-Barand ela} F.,  {Aumont} J.,  {Baccigalupi} C.,    et~al., 2014,
  \aap, 571, A11

\bibitem[\protect\citeauthoryear{{Planck Collaboration}, {Ade}, {Aghanim},
  {Alves}, {Aniano}, {Arnaud}, {Ashdown}, {Aumont}, {Baccigalupi}, {Banday}
  et~al.,}{{Planck Collaboration} et~al.}{2016}]{planck16a}
{Planck Collaboration} {Ade} P.~A.~R.,  {Aghanim} N.,  {Alves} M.~I.~R.,
  {Aniano} G.,  {Arnaud} M.,  {Ashdown} M.,  {Aumont} J.,  {Baccigalupi} C.,
  {Banday} A.~J.,    et~al., 2016, \aap, 586, A132

\bibitem[\protect\citeauthoryear{{Planck Collaboration}, {Aghanim}, {Ashdown},
  {Aumont}, {Baccigalupi}, {Ballardini}, {Band ay}, {Barreiro}, {Bartolo}
  et~al.,}{{Planck Collaboration} et~al.}{2016}]{planck16b}
{Planck Collaboration} {Aghanim} N.,  {Ashdown} M.,  {Aumont} J.,
  {Baccigalupi} C.,  {Ballardini} M.,  {Band ay} A.~J.,  {Barreiro} R.~B.,
  {Bartolo} N.,    et~al., 2016, \aap, 596, A109

\bibitem[\protect\citeauthoryear{{Radburn-Smith}, {Lucey}, {Woudt},
  {Kraan-Korteweg} \& {Watson}}{{Radburn-Smith}
  et~al.}{2006}]{2006MNRAS.369.1131R}
{Radburn-Smith} D.~J.,  {Lucey} J.~R.,  {Woudt} P.~A.,  {Kraan-Korteweg} R.~C.,
     {Watson} F.~G.,  2006, \mnras, 369, 1131

\bibitem[\protect\citeauthoryear{{Reif}, {Mebold}, {Goss}, {van Woerden} \&
  {Siegman}}{{Reif} et~al.}{1982}]{1982AuAS...50..451R}
{Reif} K.,  {Mebold} U.,  {Goss} W.~M.,  {van Woerden} H.,    {Siegman} B.,
  1982, \aaps, 50, 451

\bibitem[\protect\citeauthoryear{{Richter} \& {Huchtmeier}}{{Richter} \&
  {Huchtmeier}}{1991}]{1991AuAS...87..425R}
{Richter} O.-G.,  {Huchtmeier} W.~K.,  1991, \aaps, 87, 425

\bibitem[\protect\citeauthoryear{{Roman}, {Takeuchi}, {Nakanishi} \&
  {Saito}}{{Roman} et~al.}{1998}]{1998PASJ...50...47R}
{Roman} A.~T.,  {Takeuchi} T.~T.,  {Nakanishi} K.,    {Saito} M.,  1998, \pasj,
  50, 47

\bibitem[\protect\citeauthoryear{{Rubin}, {Ford} W.~K., {Thonnard}, {Roberts}
  \& {Graham}}{{Rubin} et~al.}{1976}]{1976AJ.....81..687R}
{Rubin} V.~C.,  {Ford} W.~K. J.,  {Thonnard} N.,  {Roberts} M.~S.,    {Graham}
  J.~A.,  1976, \aj, 81, 687

\bibitem[\protect\citeauthoryear{{Sabbadin}, {Cappellaro}, {Salvadori} \&
  {Turatto}}{{Sabbadin} et~al.}{1989}]{1989ApJ...347L...5S}
{Sabbadin} F.,  {Cappellaro} E.,  {Salvadori} L.,    {Turatto} M.,  1989,
  \apjl, 347, L5

\bibitem[\protect\citeauthoryear{{Sadler}}{{Sadler}}{1984}]{1984AJ.....89...23S}
{Sadler} E.~M.,  1984, \aj, 89, 23

\bibitem[\protect\citeauthoryear{{Sanchez-Barrantes}, {Henning}, {McIntyre},
  {Momjian}, {Minchin}, {Rosenberg}, {Schneider}, {Staveley-Smith}, {van
  Driel}, {Ramatsoku}, {Butcher} \& {Vaez}}{{Sanchez-Barrantes}
  et~al.}{2019}]{2019AJ....158..234S}
{Sanchez-Barrantes} M.,  {Henning} P.~A.,  {McIntyre} T.,  {Momjian} E.,
  {Minchin} R.,  {Rosenberg} J.~L.,  {Schneider} S.,  {Staveley-Smith} L.,
  {van Driel} W.,  {Ramatsoku} M.,  {Butcher} Z.,    {Vaez} E.,  2019, \aj,
  158, 234

\bibitem[\protect\citeauthoryear{{Sandage}}{{Sandage}}{1976}]{1976PASP...88..367S}
{Sandage} A.,  1976, \pasp, 88, 367

\bibitem[\protect\citeauthoryear{{Sanders}, {Egami}, {Lipari}, {Mirabel} \&
  {Soifer}}{{Sanders} et~al.}{1995}]{1995AJ....110.1993S}
{Sanders} D.~B.,  {Egami} E.,  {Lipari} S.,  {Mirabel} I.~F.,    {Soifer}
  B.~T.,  1995, \aj, 110, 1993

\bibitem[\protect\citeauthoryear{{Saunders}, {Sutherland}, {Maddox}, {Keeble},
  {Oliver}, {Rowan-Robinson}, {McMahon}, {Efstathiou}, {Tadros}, {White},
  {Frenk}, {Carrami{\~n}ana} \& {Hawkins}}{{Saunders}
  et~al.}{2000}]{2000MNRAS.317...55S}
{Saunders} W.,  {Sutherland} W.~J.,  {Maddox} S.~J.,  {Keeble} O.,  {Oliver}
  S.~J.,  {Rowan-Robinson} M.,  {McMahon} R.~G.,  {Efstathiou} G.~P.,  {Tadros}
  H.,  {White} S.~D.~M.,  {Frenk} C.~S.,  {Carrami{\~n}ana} A.,    {Hawkins}
  M.~R.~S.,  2000, \mnras, 317, 55

\bibitem[\protect\citeauthoryear{{Schlafly} \& {Finkbeiner}}{{Schlafly} \&
  {Finkbeiner}}{2011}]{schlafly11}
{Schlafly} E.~F.,  {Finkbeiner} D.~P.,  2011, \apj, 737, 103

\bibitem[\protect\citeauthoryear{{Schlegel}, {Finkbeiner} \&
  {Davis}}{{Schlegel} et~al.}{1998}]{schlegel98}
{Schlegel} D.~J.,  {Finkbeiner} D.~P.,    {Davis} M.,  1998, ApJ, 500, 525

\bibitem[\protect\citeauthoryear{{Schneider}, {Thuan}, {Mangum} \&
  {Miller}}{{Schneider} et~al.}{1992}]{1992ApJS...81....5S}
{Schneider} S.~E.,  {Thuan} T.~X.,  {Mangum} J.~G.,    {Miller} J.,  1992,
  \apjs, 81, 5

\bibitem[\protect\citeauthoryear{{Schr{\"o}der}, {Fl{\"o}er}, {Winkel} \&
  {Kerp}}{{Schr{\"o}der} et~al.}{2019}]{2019MNRAS.489.2907S}
{Schr{\"o}der} A.~C.,  {Fl{\"o}er} L.,  {Winkel} B.,    {Kerp} J.,  2019,
  \mnras, 489, 2907

\bibitem[\protect\citeauthoryear{{Schr{\"o}der}, {Kraan-Korteweg} \&
  {Henning}}{{Schr{\"o}der} et~al.}{2009}]{2009AuA...505.1049S}
{Schr{\"o}der} A.~C.,  {Kraan-Korteweg} R.~C.,    {Henning} P.~A.,  2009, \aap,
  505, 1049

\bibitem[\protect\citeauthoryear{{Schr{\"o}der}, {Mamon}, {Kraan-Korteweg} \&
  {Woudt}}{{Schr{\"o}der} et~al.}{2007}]{schroeder07}
{Schr{\"o}der} A.~C.,  {Mamon} G.~A.,  {Kraan-Korteweg} R.~C.,    {Woudt}
  P.~A.,  2007, \aap, 466, 481

\bibitem[\protect\citeauthoryear{{Schr{\"o}der}, {van Driel} \&
  {Kraan-Korteweg}}{{Schr{\"o}der} et~al.}{2019}]{schroeder19}
{Schr{\"o}der} A.~C.,  {van Driel} W.,    {Kraan-Korteweg} R.~C.,  2019,
  \mnras, 482, 5167

\bibitem[\protect\citeauthoryear{{Seeberger}, {Huchtmeier} \&
  {Weinberger}}{{Seeberger} et~al.}{1994}]{1994AuA...286...17S}
{Seeberger} R.,  {Huchtmeier} W.~K.,    {Weinberger} R.,  1994, \aap, 286, 17

\bibitem[\protect\citeauthoryear{{Seeberger} \& {Saurer}}{{Seeberger} \&
  {Saurer}}{1998}]{1998AuAS..127..101S}
{Seeberger} R.,  {Saurer} W.,  1998, \aaps, 127, 101

\bibitem[\protect\citeauthoryear{{Sekiguchi} \& {Wolstencroft}}{{Sekiguchi} \&
  {Wolstencroft}}{1992}]{1992MNRAS.255..581S}
{Sekiguchi} K.,  {Wolstencroft} R.~D.,  1992, \mnras, 255, 581

\bibitem[\protect\citeauthoryear{{Skrutskie}, {Cutri}, {Stiening}, {Weinberg},
  {Schneider}, {Carpenter}, {Beichman}, {Capps}, {Chester}, {Elias}
  et~al.,}{{Skrutskie} et~al.}{2006}]{2006AJ....131.1163S}
{Skrutskie} M.~F.,  {Cutri} R.~M.,  {Stiening} R.,  {Weinberg} M.~D.,
  {Schneider} S.,  {Carpenter} J.~M.,  {Beichman} C.,  {Capps} R.,  {Chester}
  T.,  {Elias} J.,    et~al., 2006, \aj, 131, 1163

\bibitem[\protect\citeauthoryear{{Spinrad}}{{Spinrad}}{1975}]{1975ApJ...199L...1S}
{Spinrad} H.,  1975, \apjl, 199, L1

\bibitem[\protect\citeauthoryear{{Springob}, {Haynes}, {Giovanelli} \&
  {Kent}}{{Springob} et~al.}{2005}]{2005ApJS..160..149S}
{Springob} C.~M.,  {Haynes} M.~P.,  {Giovanelli} R.,    {Kent} B.~R.,  2005,
  \apjs, 160, 149

\bibitem[\protect\citeauthoryear{{Staveley-Smith}, {Kraan-Korteweg},
  {Schr{\"o}der}, {Henning}, {Koribalski}, {Stewart} \&
  {Heald}}{{Staveley-Smith} et~al.}{2016}]{2016AJ....151...52S}
{Staveley-Smith} L.,  {Kraan-Korteweg} R.~C.,  {Schr{\"o}der} A.~C.,  {Henning}
  P.~A.,  {Koribalski} B.~S.,  {Stewart} I.~M.,    {Heald} G.,  2016, \aj, 151,
  52

\bibitem[\protect\citeauthoryear{{Stickel}, {Kuehr} \& {Fried}}{{Stickel}
  et~al.}{1993}]{1993AuAS...97..483S}
{Stickel} M.,  {Kuehr} H.,    {Fried} J.~W.,  1993, \aaps, 97, 483

\bibitem[\protect\citeauthoryear{{Stickel}, {Lemke}, {Klaas}, {Krause} \&
  {Egner}}{{Stickel} et~al.}{2004}]{2004AuA...422...39S}
{Stickel} M.,  {Lemke} D.,  {Klaas} U.,  {Krause} O.,    {Egner} S.,  2004,
  \aap, 422, 39

\bibitem[\protect\citeauthoryear{{Strauss}, {Huchra}, {Davis}, {Yahil},
  {Fisher} \& {Tonry}}{{Strauss} et~al.}{1992}]{1992ApJS...83...29S}
{Strauss} M.~A.,  {Huchra} J.~P.,  {Davis} M.,  {Yahil} A.,  {Fisher} K.~B.,
  {Tonry} J.,  1992, \apjs, 83, 29

\bibitem[\protect\citeauthoryear{{Sulentic} \& {Arp}}{{Sulentic} \&
  {Arp}}{1983}]{1983AJ.....88..489S}
{Sulentic} J.~W.,  {Arp} H.,  1983, \aj, 88, 489

\bibitem[\protect\citeauthoryear{{Takata}, {Yamada}, {Saito}, {Chamaraux} \&
  {Kazes}}{{Takata} et~al.}{1994}]{1994AuAS..104..529T}
{Takata} T.,  {Yamada} T.,  {Saito} M.,  {Chamaraux} P.,    {Kazes} I.,  1994,
  \aaps, 104, 529

\bibitem[\protect\citeauthoryear{{Theureau}, {Bottinelli}, {Coudreau-Durand},
  {Gouguenheim}, {Hallet}, {Loulergue}, {Paturel} \& {Teerikorpi}}{{Theureau}
  et~al.}{1998}]{1998AuAS..130..333T}
{Theureau} G.,  {Bottinelli} L.,  {Coudreau-Durand} N.,  {Gouguenheim} L.,
  {Hallet} N.,  {Loulergue} M.,  {Paturel} G.,    {Teerikorpi} P.,  1998,
  \aaps, 130, 333

\bibitem[\protect\citeauthoryear{{Theureau}, {Coudreau}, {Hallet}, {Hanski},
  {Alsac}, {Bottinelli}, {Gouguenheim}, {Martin} \& {Paturel}}{{Theureau}
  et~al.}{2005}]{2005AuA...430..373T}
{Theureau} G.,  {Coudreau} N.,  {Hallet} N.,  {Hanski} M.,  {Alsac} L.,
  {Bottinelli} L.,  {Gouguenheim} L.,  {Martin} J.~M.,    {Paturel} G.,  2005,
  \aap, 430, 373

\bibitem[\protect\citeauthoryear{{Theureau}, {Coudreau}, {Hallet}, {Hanski} \&
  {Poulain}}{{Theureau} et~al.}{2017}]{2017AuA...599A.104T}
{Theureau} G.,  {Coudreau} N.,  {Hallet} N.,  {Hanski} M.~O.,    {Poulain} M.,
  2017, \aap, 599, A104

\bibitem[\protect\citeauthoryear{{Theureau}, {Hanski}, {Coudreau}, {Hallet} \&
  {Martin}}{{Theureau} et~al.}{2007}]{2007AuA...465...71T}
{Theureau} G.,  {Hanski} M.~O.,  {Coudreau} N.,  {Hallet} N.,    {Martin}
  J.~M.,  2007, \aap, 465, 71

\bibitem[\protect\citeauthoryear{{Tifft} \& {Cocke}}{{Tifft} \&
  {Cocke}}{1988}]{1988ApJS...67....1T}
{Tifft} W.~G.,  {Cocke} W.~J.,  1988, \apjs, 67, 1

\bibitem[\protect\citeauthoryear{{Tobin}, {Hartmann}, {F{\H{u}}r{\'e}sz}, {Hsu}
  \& {Mateo}}{{Tobin} et~al.}{2015}]{2015AJ....149..119T}
{Tobin} J.~J.,  {Hartmann} L.,  {F{\H{u}}r{\'e}sz} G.,  {Hsu} W.-H.,    {Mateo}
  M.,  2015, \aj, 149, 119

\bibitem[\protect\citeauthoryear{{Tully} \& {Fisher}}{{Tully} \&
  {Fisher}}{1988}]{1988ang..book.....T}
{Tully} R.~B.,  {Fisher} J.~R.,  1988, {Catalog of Nearby Galaxies}

\bibitem[\protect\citeauthoryear{{Turatto}, {Cappellaro}, {Sabbadin} \&
  {Salvadori}}{{Turatto} et~al.}{1993}]{1993AJ....105..142T}
{Turatto} M.,  {Cappellaro} E.,  {Sabbadin} F.,    {Salvadori} L.,  1993, \aj,
  105, 142

\bibitem[\protect\citeauthoryear{{van den Bosch}, {Gebhardt}, {G{\"u}ltekin},
  {Y{\i}ld{\i}r{\i}m} \& {Walsh}}{{van den Bosch}
  et~al.}{2015}]{2015ApJS..218...10V}
{van den Bosch} R. C.~E.,  {Gebhardt} K.,  {G{\"u}ltekin} K.,
  {Y{\i}ld{\i}r{\i}m} A.,    {Walsh} J.~L.,  2015, \apjs, 218, 10

\bibitem[\protect\citeauthoryear{{van Driel}, {Schneider}, {Kraan-Korteweg} \&
  {Monnier Ragaigne}}{{van Driel} et~al.}{2009}]{vandriel09}
{van Driel} W.,  {Schneider} S.~E.,  {Kraan-Korteweg} R.~C.,    {Monnier
  Ragaigne} D.,  2009, A\&A, 505, 29

\bibitem[\protect\citeauthoryear{{Visvanathan} \& {van den
  Bergh}}{{Visvanathan} \& {van den Bergh}}{1992}]{1992AJ....103.1057V}
{Visvanathan} N.,  {van den Bergh} S.,  1992, \aj, 103, 1057

\bibitem[\protect\citeauthoryear{{Visvanathan} \& {Yamada}}{{Visvanathan} \&
  {Yamada}}{1996}]{1996ApJS..107..521V}
{Visvanathan} N.,  {Yamada} T.,  1996, \apjs, 107, 521

\bibitem[\protect\citeauthoryear{{Wegner}, {Bernardi}, {Willmer}, {da Costa},
  {Alonso}, {Pellegrini}, {Maia}, {Chaves} \& {Rit{\'e}}}{{Wegner}
  et~al.}{2003}]{2003AJ....126.2268W}
{Wegner} G.,  {Bernardi} M.,  {Willmer} C.~N.~A.,  {da Costa} L.~N.,  {Alonso}
  M.~V.,  {Pellegrini} P.~S.,  {Maia} M.~A.~G.,  {Chaves} O.~L.,    {Rit{\'e}}
  C.,  2003, \aj, 126, 2268

\bibitem[\protect\citeauthoryear{{West}, {Surdej}, {Schuster}, {Muller},
  {Laustsen} \& {Borchkhadze}}{{West} et~al.}{1981}]{1981AuAS...46...57W}
{West} R.~M.,  {Surdej} J.,  {Schuster} H.~E.,  {Muller} A.~B.,  {Laustsen} S.,
     {Borchkhadze} T.~M.,  1981, \aaps, 46, 57

\bibitem[\protect\citeauthoryear{{West} \& {Tarenghi}}{{West} \&
  {Tarenghi}}{1989}]{1989AuA...223...61W}
{West} R.~M.,  {Tarenghi} M.,  1989, \aap, 223, 61

\bibitem[\protect\citeauthoryear{{Whiteoak} \& {Gardner}}{{Whiteoak} \&
  {Gardner}}{1977}]{1977AuJPh..30..187W}
{Whiteoak} J.~B.,  {Gardner} F.~F.,  1977, Australian Journal of Physics, 30,
  187

\bibitem[\protect\citeauthoryear{{Wilson}}{{Wilson}}{1953}]{1953GCRV..C......0W}
{Wilson} R.~E.,  1953, Carnegie Institute Washington D.C. Publication, p.~0

\bibitem[\protect\citeauthoryear{{Wong}, {Ryan-Weber}, {Garcia-Appadoo},
  {Webster}, {Staveley-Smith}, {Zwaan}, {Meyer}, {Barnes}, {Kilborn}
  et~al.,}{{Wong} et~al.}{2006}]{2006MNRAS.371.1855W}
{Wong} O.~I.,  {Ryan-Weber} E.~V.,  {Garcia-Appadoo} D.~A.,  {Webster} R.~L.,
  {Staveley-Smith} L.,  {Zwaan} M.~A.,  {Meyer} M.~J.,  {Barnes} D.~G.,
  {Kilborn} V.~A.,    et~al., 2006, \mnras, 371, 1855

\bibitem[\protect\citeauthoryear{{Woudt}, {Kraan-Korteweg}, {Cayatte},
  {Balkowski} \& {Felenbok}}{{Woudt} et~al.}{2004}]{2004AuA...415....9W}
{Woudt} P.~A.,  {Kraan-Korteweg} R.~C.,  {Cayatte} V.,  {Balkowski} C.,
  {Felenbok} P.,  2004, \aap, 415, 9

\bibitem[\protect\citeauthoryear{{Woudt}, {Kraan-Korteweg} \&
  {Fairall}}{{Woudt} et~al.}{1999}]{1999AuA...352...39W}
{Woudt} P.~A.,  {Kraan-Korteweg} R.~C.,    {Fairall} A.~P.,  1999, \aap, 352,
  39

\bibitem[\protect\citeauthoryear{{Woudt}, {Kraan-Korteweg}, {Lucey}, {Fairall}
  \& {Moore}}{{Woudt} et~al.}{2008}]{2008MNRAS.383..445W}
{Woudt} P.~A.,  {Kraan-Korteweg} R.~C.,  {Lucey} J.,  {Fairall} A.~P.,
  {Moore} S.~A.~W.,  2008, \mnras, 383, 445

\bibitem[\protect\citeauthoryear{{Wouterloot} \& {Brand}}{{Wouterloot} \&
  {Brand}}{1989}]{1989AuAS...80..149W}
{Wouterloot} J.~G.~A.,  {Brand} J.,  1989, \aaps, 80, 149

\bibitem[\protect\citeauthoryear{{Wynn-Williams}, {Heasley}, {Depoy}, {Hill} \&
  {Becklin}}{{Wynn-Williams} et~al.}{1986}]{1986ApJ...304..409W}
{Wynn-Williams} C.~G.,  {Heasley} J.~N.,  {Depoy} D.~L.,  {Hill} G.~J.,
  {Becklin} E.~E.,  1986, \apj, 304, 409

\bibitem[\protect\citeauthoryear{{Yamada} \& {Saito}}{{Yamada} \&
  {Saito}}{1993}]{1993PASJ...45...25Y}
{Yamada} T.,  {Saito} M.,  1993, \pasj, 45, 25

\bibitem[\protect\citeauthoryear{{Yamada}, {Tomita}, {Saito}, {Chamaraux} \&
  {Kazes}}{{Yamada} et~al.}{1994}]{1994MNRAS.270...93Y}
{Yamada} T.,  {Tomita} A.,  {Saito} M.,  {Chamaraux} P.,    {Kazes} I.,  1994,
  \mnras, 270, 93

\end{thebibliography}



\appendix

\section{Catalogue of velocity measurements and k-corrections}

We searched the online databases HyperLeda, NED and Simbad for velocity
information on our objects. In addition, we searched particular
publications with the CDS portal for more comprehensive
information. Finally, we supplemented the velocity database with the latest
data from our on-going observation programmes in the optical and \HI , to
be published presently. For each measurement that we found we determined
whether it was an optical or \HI\ measurement, and whether the reference
was original or a compilation. If more than one velocity measurement was
found, we chose the best of the optical and the best of the
\HI\ measurements where `best' was usually determined by the error and
sometimes by additional information like the velocity resolution of an
\HI\ observation. No detailed comparison has been attempted. For our final
catalogue, we usually preferred an \HI\ measurement over an optical for its
greater accuracy.

The catalogue of the adopted velocity measurements and k-corrections is
available online, where Table~\ref{kctabex} lists the first 10 galaxy
entries. The table is in the same format as the full object catalogue
presented in Paper I with the same ID per row. Some columns are repeated
for easy reference. The references for the bibliographic code information
used are listed in Table~\ref{refvelextab} (the full table is available
online). The columns of Table~\ref{kctabex} are as follows:

\begin{description}

\item{\it Col.\ 1:} ID: 2MASX catalogue identification number (based on
  J2000.0 coordinates);

\item{\it Col.\ 2a and 2b:} Galactic coordinates: longitude $l$ and
  latitude $b$, in degrees;

\item{\it Col.\ 3:} Extinction: \ebv\ value derived from the DIRBE/IRAS
  maps (SFD), in mag;

\item{\it Col.\ 4:} Colour \jko\ corrected for foreground extinction;
  contrary to the values given in Paper I, these were corrected with the
  3-digit precision extinction coefficients presented in
  Table~\ref{elawtab}, in mag;

\item{\it Col.\ 5:} Sample flag galaxy: `g' denotes a galaxy (object
  classes $1-4$), and `p' stands for galaxy candidates (class 5);

\item{\it Col.\ 6:} Object offset flag: `o' stands for coordinates that are
  offset from the centre of the object, and `e' stands for detections near
  the edge of an image (at an offset position from the object centre) which
  were detected with properly centred coordinates on the adjacent image;

\item{\it Col.\ 7:} Velocity flag for the final adopted velocity: o' stands
  for an optical measurement and `o?' indicates a questionable optical
  measurement or object ID, `g' the velocity shows it to be a Galactic
  object, `h' stands for published \HI\ velocity and `h:' for uncertain
  \HI\ velocity, `h?' is an uncertain object ID (based on a radio
  telescope's large beam size that may include nearby galaxies), and a star
  indicates a yet unpublished measurement;


\item{\it Col.\ 8:} 2MRS flag: a star denotes if our adopted velocity,
  error or reference deviates from the values given in the 2MRS catalogue
  (\citealt{huchra12}, \citealt{macri19});

\item{\it Col.\ 9a and 9b:} Adopted velocity and error according to the
  flag in Col.\ 7, in \kms ; a star replaces the values in case of
  unpublished values;

\item{\it Col.\ 10:} k-correction for the colour \jk , in mag;

\item{\it Col.\ 11a -- 11c:} Adopted optical velocity with error and
  reference (bibliographic code) for this object, in \kms ; square brackets
  indicate a discarded measurement; a star replaces the values in case of
  unpublished measurements;

\item{\it Col.\ 12a -- 12c:} Adopted \HI\ velocity with error and reference
  (bibliographic code) for this object, in \kms ; square brackets indicate
  a discarded measurement; a star replaces the values in case of
  unpublished measurements.

\end{description}

The table lists also some velocity information as yet unpublished. These
are identified with a star and are from our observing campaigns: Optical
observations in the frame work of the 2MRS/2MZOA surveys are identified by
observation year, the ID 2MRS or 2MZO and the observatory ID as explained
in \citet{macri19}. \HI\ observations obtained by our group refer to the
blind Parkes \HI\ Galactic bulge survey (Kraan-Korteweg \etal , in prep.)
and to pointed \HI\ observations of or sample with the Parkes telescope
(Said, priv.comm.) and the NRT (Kraan-Korteweg \etal , in prep.). Finally,
some velocities quoted by Simbad do not have a reference.

\onecolumn 
{\tiny 
\begin{longtable}{@{\extracolsep{1.0mm}}l@{\extracolsep{2.0mm}}r@{\extracolsep{1.0mm}}r@{\extracolsep{2.0mm}}r@{\extracolsep{1.0mm}}r@{\extracolsep{1.0mm}}c@{\extracolsep{1.0mm}}c@{\extracolsep{1.0mm}}l@{\extracolsep{1.0mm}}c@{\extracolsep{1.0mm}}r@{\extracolsep{1.0mm}}r@{\extracolsep{1.0mm}}r@{\extracolsep{2.0mm}}r@{\extracolsep{1.0mm}}r@{\extracolsep{2.0mm}}l@{\extracolsep{1.0mm}}r@{\extracolsep{1.0mm}}r@{\extracolsep{2.0mm}}l}
\caption{{2MZOA catalogue with k-correction; example page, the full table is
    available online} \label{kctabex}}\\
\hline
\noalign{\smallskip}
2MASX J          &  $l$ &  $b$ & EBV & ($J$-$K)^o$ & \multicolumn{4}{c}{Flags} & $v$(adpt) & $v_{\rm err}$ & k-corr & $v$(op) & $v_{\rm err}$ &  Ref  & $v$(HI)  & $v_{\rm err}$ &  Ref \\
                 &  deg &  deg & mag &    mag    & gal & off & vel & 2MRS    & km/s        & km/s          &  mag   & km/s          & km/s          &       & km/s          & km/s          &      \\
(1)              & (2a) & (2b) & (3) &    (4)    & (5) & (6) & (7) & (8)     & (9a)        & (9b)          & (10)   & (11a)         & (11b)         & (11c) & (12a)         & (12b)         & (12c) \\
\noalign{\smallskip}
\hline
\noalign{\smallskip}
\endfirsthead
\caption{continued.}\\
\hline
\noalign{\smallskip}
2MASX J          &  $l$ &  $b$ & EBV & ($J$-$K)^o$ & \multicolumn{4}{c}{Flags} & $v$(adpt) & $v_{\rm err}$ & k-corr & $v$(op) & $v_{\rm err}$ &  Ref  & $v$(HI)  & $v_{\rm err}$ &  Ref \\
                 &  deg &  deg & mag &    mag    & gal & off & vel & 2MRS    & km/s         & km/s          &  mag   & km/s          & km/s          &       & km/s          & km/s          &      \\
(1)              & (2a) & (2b) & (3) &    (4)    & (5) & (6) & (7) & (8)     & (9a)         & (9b)          & (10)   & (11a)         & (11b)         & (11c) & (12a)         & (12b)         & (12c) \\
\noalign{\smallskip}
\hline
\noalign{\smallskip}
\endhead
\noalign{\smallskip}
\hline
\endfoot
00000637$+$5319136 & 115.217 &$ -8.780$ &$  0.300$ &  0.99 & g& $-$& o &$-$&  12447 & 65 &  0.05 &     12447 & 65 &2012ApJS..199...26H  &    $-$ & $-$&$-$                 \\
00031331$+$5352149 & 115.783 &$ -8.330$ &$  0.299$ &  1.05 & g& $-$& o &$-$&  11772 & 30 &  0.05 &     11772 & 30 &2012ApJS..199...26H  &    $-$ & $-$&$-$                 \\
00033726$+$6919064 & 118.707 &$  6.840$ &$  0.808$ &  1.12 & g& $-$& o &$-$&   6864 &105 &  0.03 &      6864 &105 &2012ApJS..199...26H  &    $-$ & $-$&$-$                 \\
00034142$+$7036434 & 118.955 &$  8.110$ &$  0.789$ &  0.87 & g& $-$& o &$-$&   4620 & 16 &  0.02 &      4620 & 16 &2012ApJS..199...26H  &    $-$ & $-$&$-$                 \\
00053428$+$5337528 & 116.084 &$ -8.629$ &$  0.417$ &  1.01 & g& $-$& h &$-$&   9523 &  7 &  0.04 &       $-$ & $-$&$-$                  &   9523 &  7 &1998A\&AS..130..333T \\
00083102$+$5301491 & 116.417 &$ -9.298$ &$  0.411$ &  1.02 & g& $-$& o &$-$&  12124 & 38 &  0.05 &     12124 & 38 &2012ApJS..199...26H  &    $-$ & $-$&$-$                 \\
00083403$+$5530408 & 116.844 &$ -6.853$ &$  0.432$ &  1.05 & g& $-$& h &$-$&   5155 &  4 &  0.02 &       $-$ & $-$&$-$                  &   5155 &  4 &2003A\&A...412...57P \\
00103893$+$5325040 & 116.800 &$ -8.968$ &$  0.265$ &  1.00 & g& $-$& o &$-$&   9900 & 23 &  0.04 &      9900 & 23 &2012ApJS..199...26H  &    $-$ & $-$&$-$                 \\
00141253$+$7036448 & 119.824 &$  7.966$ &$  1.014$ &  0.92 & g& $-$& o & * &   6958 &  8 &  0.03 &      6973 & 43 &2012ApJS..199...26H  &   6958 &  8 &2018MNRAS.481.1262K \\
00161976$+$7025219 & 119.974 &$  7.753$ &$  1.046$ &  0.43 & g& $o$& o*&$-$&      * &  * &  0.15 &         * &  * &20182MZO.FLWO.0000M  &    $-$ & $-$&$-$                 \\
\end{longtable}
 
} 
{\small 
\begin{table} 
\centering
\begin{minipage}{140mm}
\caption{List of references for velocity measurements; example page, the full table is
    available online} \label{refvelextab} 
\begin{tabular}{ll}
\hline
 Bibcode             &  Reference                    \\
\hline
1998A\&AS..130..333T  &  \citet{1998AuAS..130..333T}  \\
2012ApJS..199...26H  &  \citet{2012ApJS..199...26H}  \\
2018MNRAS.481.1262K  &  \citet{2018MNRAS.481.1262K}  \\
\noalign{\smallskip}
\multicolumn{2}{l}{Unpublished optical and \HI\ observations by our team:} \\
20182MZO.FLWO.0000M  &  Macri, priv.comm.            \\
\hline
\end{tabular}
\end{minipage}
\end{table}

}
\twocolumn

\section{Variation of $f$-values across the ZoA for the Planck extinction maps}

\begin{figure*} 
\centering
\includegraphics[width=0.7\textwidth]{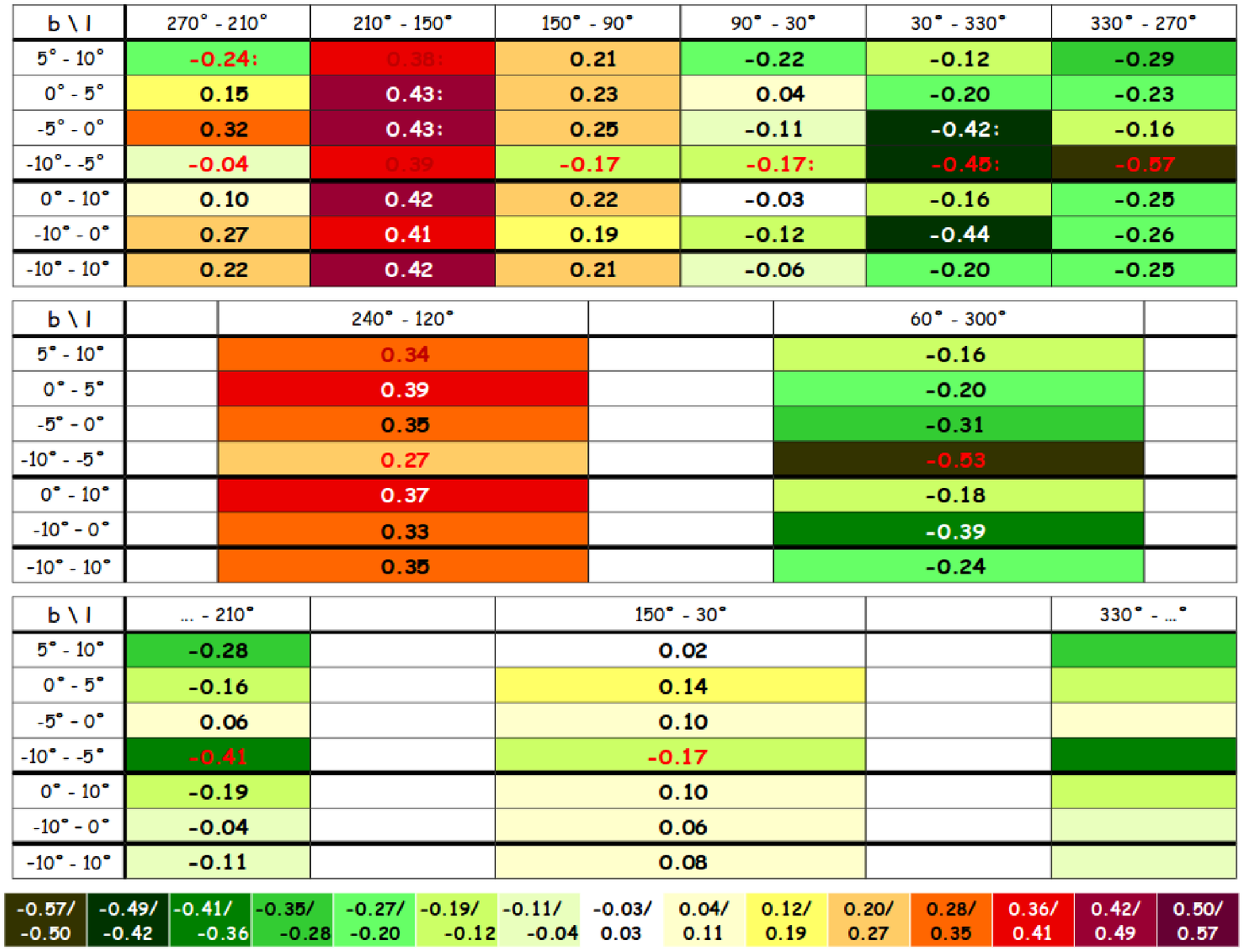}
\caption{Same as Fig.~\ref{lbmap2plot} for the P-PR1 extinctions. 
  Note the change to the colour scale.
}
\end{figure*}
\begin{figure*} 
\centering
\includegraphics[width=0.7\textwidth]{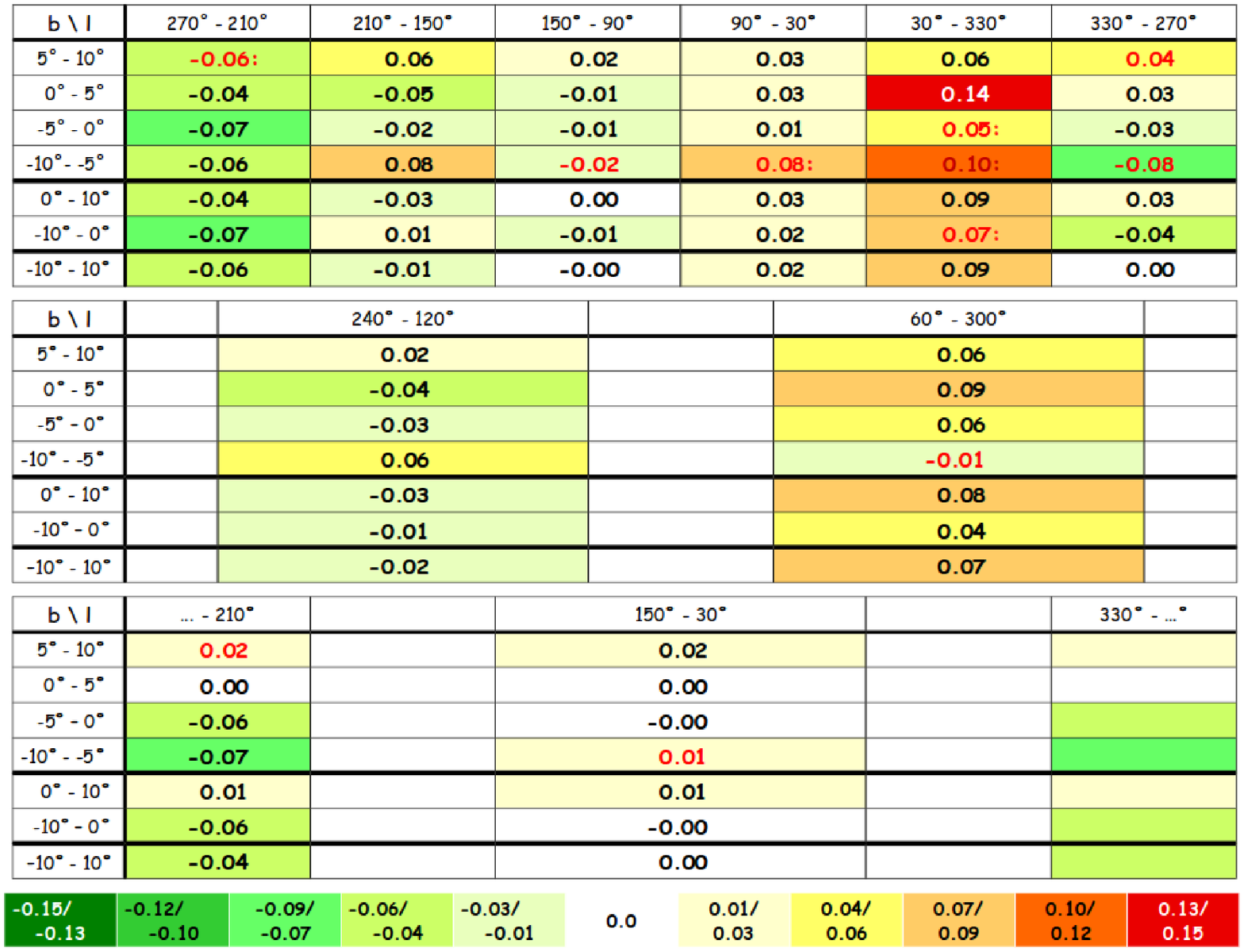}
\caption{Same as Fig.~\ref{lbmap2plot} for the P-MF extinctions. 
}
\end{figure*}
\begin{figure*} 
\centering
\includegraphics[width=0.7\textwidth]{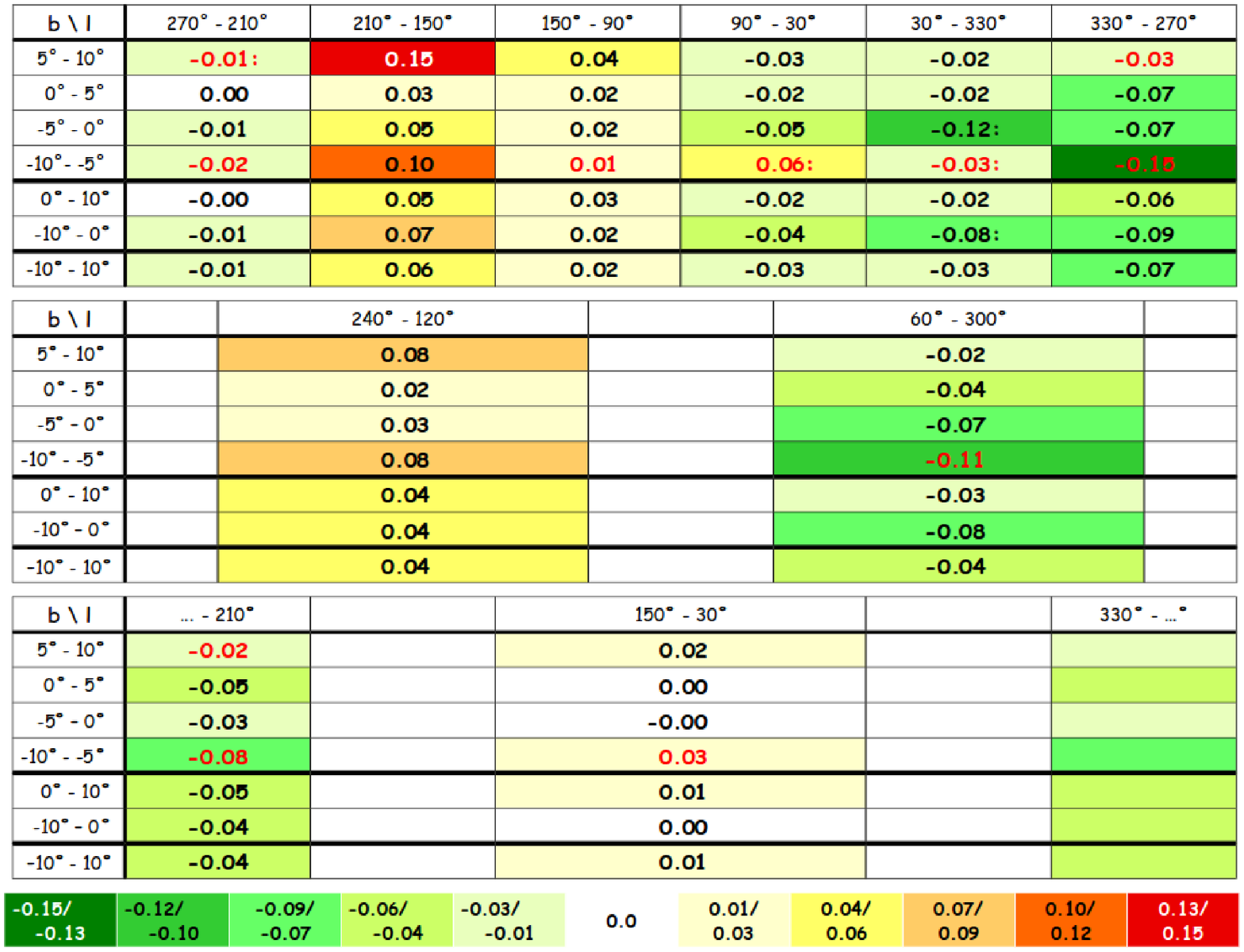}
\caption{Same as Fig.~\ref{lbmap2plot} for the P-AV extinctions. 
}
\end{figure*}
\begin{figure*} 
\centering
\includegraphics[width=0.7\textwidth]{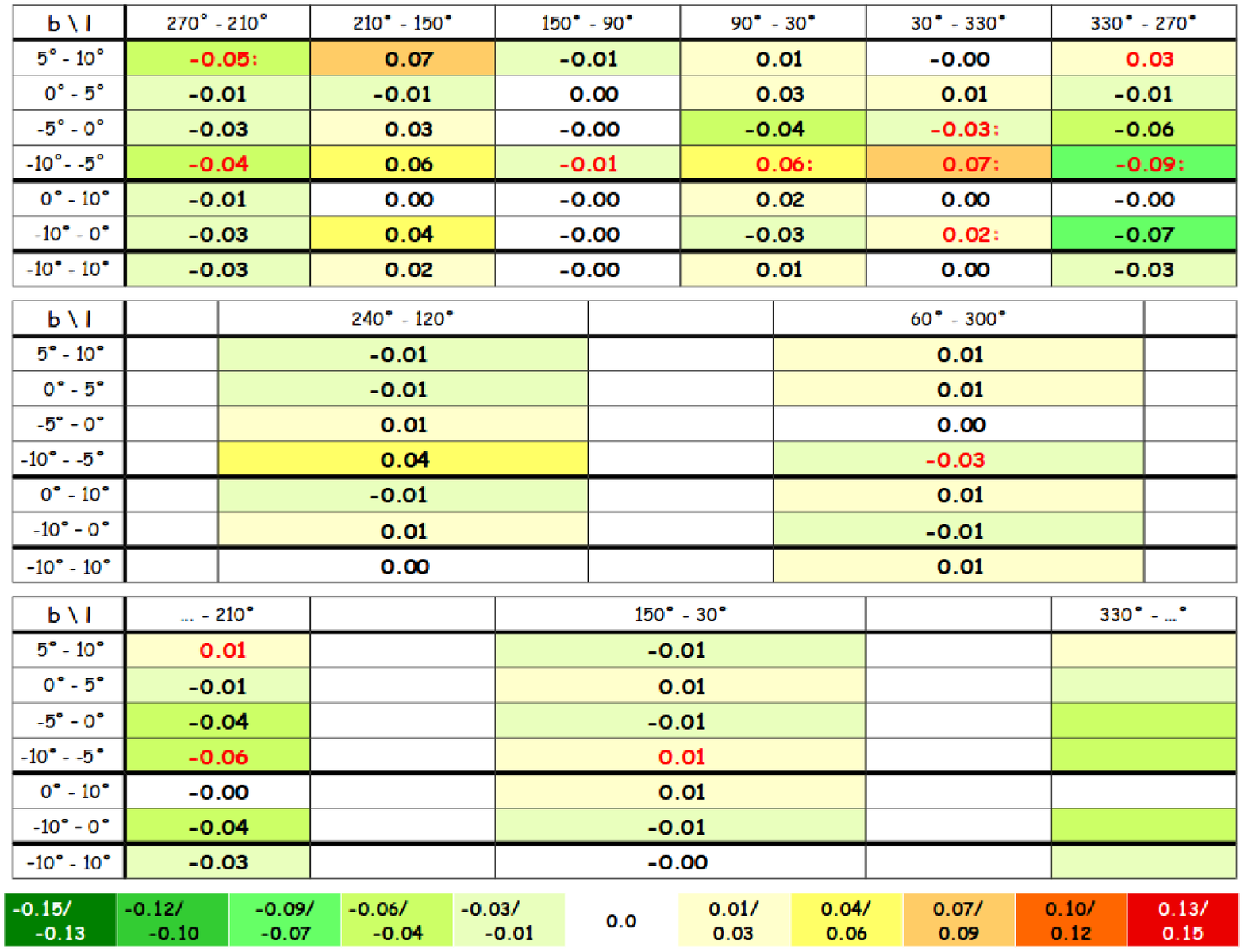}
\caption{Same as Fig.~\ref{lbmap2plot} for the P-GNILC extinctions. 
}
\end{figure*}

\bsp
\label{lastpage}
\end{document}